\documentclass[journal]{IEEEtran}
\usepackage{amsmath,amsfonts,amssymb}
\usepackage{algorithmic}
\usepackage{array}
\usepackage[caption=false,font=normalsize,labelfont=sf,textfont=sf]{subfig}
\usepackage{textcomp}
\usepackage{stfloats}
\usepackage{url}
\usepackage{xcolor}
\usepackage{verbatim}
\usepackage{graphicx}
\usepackage{adjustbox}
\usepackage{tikz}
\usepackage{bm}
\usepackage{multirow}
\usepackage{booktabs}
\usetikzlibrary{shapes, arrows, positioning, math, , decorations.pathreplacing}
\tikzstyle{arrow} = [thick,->,>=stealth]
\usetikzlibrary{chains,shapes.multipart}
\usetikzlibrary{shapes,calc}
\usetikzlibrary{automata}

\def\BibTeX{{\rm B\kern-.05em{\sc i\kern-.025em b}\kern-.08em
    T\kern-.1667em\lower.7ex\hbox{E}\kern-.125emX}}
\usepackage{balance}
\usepackage{etoolbox}
\makeatletter
\patchcmd{\@makecaption}
  {\scshape}
  {}
  {}
  {}
\makeatother

\begin{document}
\title{AC Loss Computation in Large-Scale \\ Low-Temperature Superconducting Magnets: \\ Multi-Scale and Semi-Analytical Procedures}
\author{Louis Denis, Vincent Nuttens, Benoît Vanderheyden, and Christophe Geuzaine
\thanks{L. Denis is a research fellow funded by the F.R.S-FNRS. (\textit{Corresponding author: L. Denis})

L. Denis, B. Vanderheyden, and C. Geuzaine are with the Department of Electrical Engineering and Computer Science, Institut Montefiore B28 in the University of Liege, 4000 Liege, Belgium (e-mail: louis.denis@uliege.be).

V. Nuttens is with Ion Beam Applications, 1348 Louvain-la-Neuve, Belgium.}}


\makeatletter
\def\ps@IEEEtitlepagestyle{
  \def\@oddfoot{\mycopyrightnotice}
  \def\@evenfoot{}
}
\def\mycopyrightnotice{
  {\footnotesize
  \begin{minipage}{\textwidth}
    \fbox{\parbox{\textwidth}{
        This work has been submitted to the IEEE for possible publication. Copyright may be transferred without notice, after which this version may no longer be accessible.
        }}
  \end{minipage}
  }
}

\maketitle

\begin{abstract}
In this paper, we introduce two complementary approaches for the accurate prediction of AC losses in large-scale low-temperature superconducting (LTS) magnets. These methods account for the temperature rise within the LTS coil and its impact on AC losses. The first approach is multi-scale and relies on the coupling between a macroscopic homogenized model of the LTS coil and a mesoscopic model of a single conductor for loss prediction. The second approach is semi-analytical and is based on analytical approximations for the hysteresis losses, which are validated against a single filament model. The second approach offers a faster computation suitable for initial design considerations, while the multi-scale method is shown to provide more accurate results. We apply both methods to the prediction of AC losses generated in the LTS coil inside the IBA S2C2 synchrocyclotron during its ramp-up procedure. Additionally, we discuss the convergence properties of the multi-scale approach and demonstrate the good agreement between the numerical results and experimental data.
\end{abstract}

\begin{IEEEkeywords}
AC losses, finite-elements, low-temperature superconductors, magneto-thermal analysis, multi-scale.
\end{IEEEkeywords}

\section{Introduction}
\IEEEPARstart{I}{n} transient situations such as a magnet ramp-up, superconducting (SC) wires experience heat generation due to the time variation of the local magnetic field. Predicting AC losses and the corresponding temperature rise is crucial for the design of SC magnets and their cryogenic systems.

The Low-Temperature Superconducting (LTS) technology based on Nb-Ti and Nb$_3$Sn is widely used as it is currently less expensive and more mature than the High-Temperature Superconducting~(HTS) technology. LTS represent the biggest share of the market for superconducting magnets, as they are used in medical applications such as MRI and NMR systems \cite{SuperconductingMaterialsChallenges}. With the supply of liquid helium becoming a concern \cite{LiquidHeliumConcern}, the interest in AC loss reduction in LTS magnets is growing. Innovations in medical applications are leaning towards dry magnets, which are based on conduction cooling and are therefore more sensitive to AC losses than magnets operating directly in liquid helium \cite{WaldDryMagnets}. LTS magnets also find applications beyond medical settings, including the use of Nb$_3$Sn in next-generation particle accelerators \cite{TodescoHL-LHC, AbadaFCC}, their deployment in the ITER project \cite{IterProject}, and their role in the development of hybrid superconducting magnets \cite{HybridMagnet}.

In the past decade, the finite-element (FE) method has become the reference to compute AC losses in superconducting devices \cite{Grilli2014, ReviewShen}. It constitutes a versatile tool for the magneto-thermal modelling of superconductors, as it allows convenient coupling between electromagnetic and thermal physics \cite{ShenGrilliOverviewH}. Nevertheless, the large-scale nature of superconducting magnets, combined to the non-linear behaviour of superconductors, makes the computation of AC losses demanding~\cite{ChallengesSiroisGrilli}. While domain decomposition methods \cite{RivaDDM} or parallel-in-time methods \cite{SchnaubeltParareal} can help reduce the computational effort associated with 3D simulations, the brute-force simulation of large-scale systems remains challenging as the corresponding computational resources are still prohibitive.

Currently, the prediction of AC losses in large-scale LTS magnets is thus mainly limited to analytical approximations as found in \cite{Carr, Wilson, YwasaBook}. Recent contributions, such as the helicoidal transformation method~\cite{DularHelicoidal} or the coupled axial and transverse currents (CATI) method~\cite{DularCATI}, aim to reduce the computational resources required to simulate LTS conductors. However, these methods are not yet adapted to the description of large-scale LTS magnets.

By contrast, many modelling strategies have been introduced to further reduce the computational load associated with the simulation of large-scale HTS systems. Homogenized models have been proposed for both 2D \cite{ZermenoHom2D} and 3D \cite{ZermenoHom3D} simulations. Moreover, a multi-scale method has been introduced in \cite{QuevalFirstMultiScale, Berrospe1} to simulate stacks of HTS tapes, which relies on two separate models: the \textit{coil} submodel and the \textit{single-tape} submodel. The first submodel provides the approximate background field to which the single tapes are subjected, while the second submodel computes the local AC losses as well as the tape magnetic response to be taken into account at the coil scale. An extensive description and comparison of these methods can be found in \cite{BerrospeFinal}, as they allow accurate predictions while significantly reducing computational resources. 

Building upon some concepts initially introduced by the HTS modelling community, the present study aims at providing a robust and flexible framework for the accurate AC loss computation in LTS magnets, while accounting for the corresponding temperature rise in the LTS coil. To this end, this work presents both \textit{multi-scale} and \textit{semi-analytical} magneto-thermal FE methods that allow the simulation of large-scale LTS magnets. While the described framework is general, the methods are illustrated with magnets in medical cyclotrons for hadrontherapy, in which operating conditions change over several hours. The multi-scale approach (MSA) relies on the coupling between a \textit{macroscopic} model of the LTS coil and a \textit{mesoscopic} model of a single conductor for loss prediction. Conversely, the semi-analytical approach (SAA) relies exclusively on analytical AC loss approximations. The two methods therefore provide complementary tools, with the SAA being faster yet less accurate than the MSA.

The paper is structured as follows: the numerical problem to be solved is described in Section~\ref{sec:problem-description}, in the geometry of a generic medical LTS magnet. In Section~\ref{sec:analytical-approximations}, analytical approximations for the losses in LTS magnets are reviewed and adapted, which provide a first reference for the numerical results. The mesoscopic model of a single filament is characterized in Section~\ref{sec:characterization-meso}, in which its underlying assumptions are discussed as well as its validity range. The accuracy of the analytical approximations introduced previously is assessed. The principle of the proposed MSA is described in Section~\ref{sec:multi-scale-approach}, followed by implementation details using the open-source software GetDP \cite{GetDP1, GetDP2}. The SAA is next introduced in Section~\ref{sec:semi-analytical-approach}. Finally, both methods are applied to the ramp-up of the S2C2 synchrocyclotron~\cite{S2C2} in Section~\ref{sec:results}. The convergence of the MSA is discussed, while its results are compared to predictions from the SAA as well as to experimental data. Conclusions are given in Section~\ref{sec:conclusion}.

\section{Description of the Macroscopic Problem}
\label{sec:problem-description}
\begin{figure*}[!t]
    \begin{center}
    \includegraphics{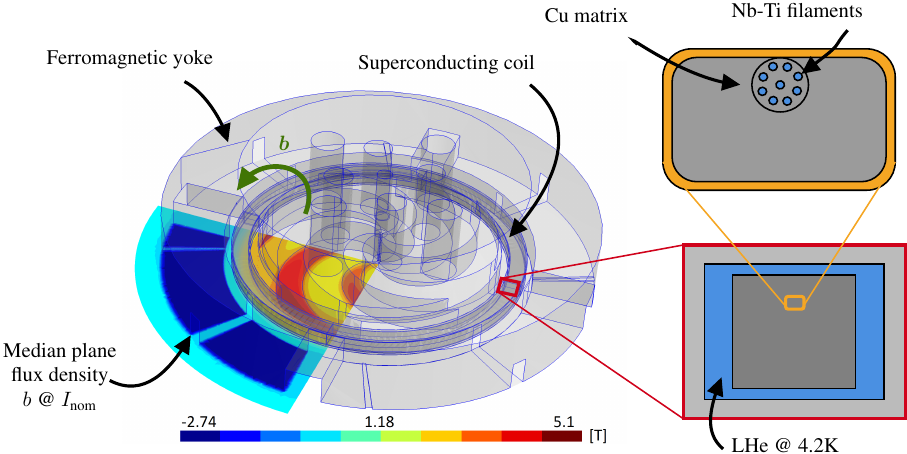}
    \end{center}
    \caption{The C400 cyclotron and the inner structure of its superconducting coil, along with the median plane normal magnetic flux density $b$~[T] in nominal current conditions. The coil is cooled down by liquid helium (LHe) at $4.2$~K. The inner structure of the coil is not to scale: it contains more than 1300 composite wire-in-channel conductor turns with much smaller and more closely arranged filaments. Yoke diameter: $ d_{\text{y}}\approx 7$~m, filament diameter: $ d_{\text{f}}\approx 50$~µm.}
    \label{fig:C400_geo}
\end{figure*}

\noindent As mentioned, the goal is to predict the AC losses occurring in large-scale LTS magnets. For the sake of illustration, LTS coils placed within medical cyclotrons are considered in this paper, such as the C400 \cite{C400} and the S2C2 developed by Ion Beam Applications (IBA). The C400, which has been studied in~\cite{C400_Denis} using a simplified version of the multi-scale approach, is depicted in Fig.~\ref{fig:C400_geo}. Such coils are made of multifilamentary LTS conductors, with the SC filaments having a transversal dimension much smaller than the size of the magnet. This is referred to as the \textit{separation of scales} illustrated in Fig.~\ref{fig:C400_geo}. The ferromagnetic yoke is supposed to have a three-dimensional geometry with a planar symmetry in the median plane. The macroscopic structure of the coil itself is assumed axisymmetric. The focus is set on the ramp-up of the magnet, which is the increase from zero current to the nominal current required to accelerate ionized particles. Typically, the ramp-up of medical cyclotrons lasts several hours (e.g. 2h for the C400 magnet and 4h for the S2C2 magnet). To account for the impact of temperature, a coupled magneto-thermal FE problem is solved at the \textit{macroscopic} scale of the magnet. The magnetic and thermal domains of interest are denoted by $\Omega_{\text{M,mag}}$ and $\Omega_{\text{M,the}}$, representing respectively the space occupied by the magnet (with its surrounding air) and the cold mass of the cyclotron, such that $\Omega_{\text{M,the}} \subset \Omega_{\text{M,mag}}$. By symmetry, only the volume above the median plane of the magnet is considered.

\subsection{Magnetic macroscopic formulation} \label{sec:macro-mag}
\noindent Maxwell's equations are solved in the magnetoquasistatic approximation \cite{JacksonTheory}:
\begin{equation}
    \textbf{curl}~\bm{e}_{\text{M}} = -\partial_t \bm{b}_{\text{M}}, \quad \textbf{curl}~\bm{h}_{\text{M}} = \bm{j}_{\text{M}}, \quad \text{and} \quad \text{div}~\bm{b}_{\text{M}} = 0,
\end{equation}
with $\bm{e}_{\text{M}}$, $\bm{b}_{\text{M}}$, $\bm{h}_{\text{M}}$, $\bm{j}_{\text{M}}$ the electric field (V/m), the magnetic flux density (T), the magnetic field (A/m), and the current density (A/m$^2$), respectively. The $\cdot_{\text{M}}$ subscript refers to the fields at the macroscopic scale. The set of equations is closed with material laws: $\bm{b}_{\text{M}} = \mu \bm{h}_{\text{M}}$ and $\bm{j}_{\text{M}} = \sigma \bm{e}_{\text{M}}$, with $\mu$, $\sigma$ the magnetic permeability (H/m) and the electrical conductivity~(S/m), respectively. The magnetic domain is composed of the conducting domain $\Omega_{\text{M,c}}$, in which $\sigma \neq 0$, and its complementary non-conducting domain $\Omega_{\text{M,c}}^{\text{C}} = \Omega_{\text{M,mag}} \setminus \Omega_{\text{M,c}}$.

In the present study, the magnetic response due to eddy currents in SC filaments is neglected at the macroscopic scale and the current density is assumed to be equal to the engineering current density $\bm{j}_{\text{M}} = \bm{j}_{\text{eng}}$ in the LTS coil $\Omega_{\text{M,s}} \subset \Omega_{\text{M,c}}^{\text{C}}$. In this context, the macroscopic problem only involves ferromagnetic materials and a modified vector potential $a$-formulation is preferred for convergence issues \cite{DularFEFormulationsSC}. The modified vector potential $\bm{a}_{\text{M}}$ is the main unknown. It is defined such that:
\begin{equation} 
    \bm{b}_{\text{M}} = \textbf{curl}~\bm{a}_{\text{M}} \text{ in } \Omega_{\text{M,mag}}, \quad \text{and} \quad \bm{e}_{\text{M}} = -\partial_t \bm{a}_{\text{M}} \text{ in } \Omega_{\text{M,c}}.
\end{equation}
In $\Omega_{\text{M,c}}^{\text{C}}$, the uniqueness of $\bm{a}_{\text{M}}$ is ensured using the co-tree gauge \cite{JDularPhd}.

The boundary of the magnetic domain is denoted by $\Gamma_{\text{M,mag}} = \partial \Omega_{\text{M,mag}} = \Gamma_{\text{M,mag},e} \cup \Gamma_{\text{M,mag},h}$, with $\bm{n}$ the outward pointing normal unit vector. A homogeneous essential boundary condition is applied at infinity: $\bm{a}_{\text{M}} \times \bm{n} = \bm{0}$ on $\Gamma_{\text{M,mag},e}$. An infinite shell transformation \cite{InfiniteShellTransfo} is used to map the unbounded domain to a numerical region of finite size. A homogeneous natural boundary condition is applied on the median plane of the magnet: $\bm{h}_{\text{M}} \times \bm{n} = \bm{0}$ on $\Gamma_{\text{M,mag},h}$.

The magnetic vector potential is approximated using edge basis functions and belongs to the space of square integrable functions with a square integrable curl in $\Omega_{\text{M,mag}}$, such that $\bm{a}_{\text{M}} \in \mathcal{H}(\text{curl},\Omega_{\text{M,mag}}) \times ]0,T_{\text{sim}} ]$, with $T_{\text{sim}}$ the final simulation time. At the macroscopic scale, the magnetodynamic weak formulation reads: \\
From an initial solution at $t=0$, find $\bm{a}_{\text{M}} \in \mathcal{H}(\text{curl},\Omega_{\text{M,mag}})\times ]0,T_{\text{sim}} ]$ s.t.,
\begin{multline}
    (\nu~\textbf{curl}~\bm{a}_{\text{M}}, \textbf{curl}~\bm{a}_{\text{M}}')_{\Omega_{\text{M,mag}}} + (\sigma~\partial_t \bm{a}_{\text{M}}, \bm{a}_{\text{M}}')_{\Omega_{\text{M,c}}} \\ = (\bm{j}_{\text{eng}}, \textbf{curl}~\bm{a}_{\text{M}}')_{\Omega_{\text{M,s}}}, \label{eq:macro_magnetic_weak_form}
\end{multline}
$\forall \bm{a}_{\text{M}}' \in \mathcal{H}(\text{curl},\Omega_{\text{M,mag}})$ with $\bm{a}_{\text{M}}' \times \bm{n} = \bm{0}$ on $\Gamma_{\text{M,mag},e}$, where $(\cdot,\cdot)_{\Omega}$ denotes the $L^2$ inner product over the domain $\Omega$. 

The time discretization of \eqref{eq:macro_magnetic_weak_form} is performed with the implicit Euler (IE) (or \textit{backward Euler}) method. As the reluctivity $\nu = \mu^{-1} = \nu(\textbf{curl}~\bm{a}_{\text{M}})$ is non-linear in the ferromagnetic yoke $\Omega_{\text{M,mag}}^{\text{NL}} \subset \Omega_{\text{M,mag}}$, an iterative Newton-Raphson (NR) scheme is performed at each time step of the macroscopic magnetic simulation. Outside of the yoke, the reluctivity is considered constant: $\nu(\bm{x}_{\text{M}}) = 1/ \mu_0 \, \forall \, \bm{x}_{\text{M}} \in \Omega_{\text{M,mag}} \setminus \Omega_{\text{M,mag}}^{\text{NL}}$.

\subsection{Thermal macroscopic formulation} \label{sec:macro-the}
\noindent In the thermal domain $\Omega_{\text{M,the}}$, the heat equation \cite{HeatBook} is solved:
\begin{equation}
    \rho c_p \partial_t T_{\text{M}} + \text{div}~ \bm{q}_{\text{M}}'' = \bar{q}_{\text{s}}, \label{eq:heat_eq}
\end{equation}
with $T_{\text{M}}$, $\bm{q}_{\text{M}}''$ and $\bar{q}_{\text{s}}$ the temperature (K), the heat flux (W/m$^2$) and the heat source (W/m$^3$), respectively. The heat flux is modelled using Fourier's law: 
\begin{equation}
    \bm{q}_{\text{M}}'' = -\bm{\kappa} \cdot \textbf{grad}~T_{\text{M}}.
\end{equation}
Material properties are the mass density $\rho$ (kg/m$^3$), the specific heat capacity $c_p$ (J/kg/K) and the thermal conductivity $\bm{\kappa}$ (W/m/K). At the macroscopic scale, the thermal properties of the composite LTS coil are homogenized using thermal resistances \cite{HeatBook} as done in Appendix A. In this context, the (anisotropic) equivalent thermal conductivity $\bm{\kappa}$ is a tensor.

The boundary of the thermal domain is denoted by $\Gamma_{\text{M,the}}$ and no essential (Dirichlet) boundary conditions are assumed. Only natural (Neumann) boundary conditions are imposed: $\bm{q}_{\text{M}}''~\cdot~\bm{n}~=~\bar{f}(T_{\text{M}})$ on $\Gamma_{\text{M,the}}$, with $\bar{f}$ representing one of the following boundary conditions:
\begin{equation}
    \bar{f}(T_{\text{M}}) =
\begin{cases}0 & \text{adiabatic surface},\\
q_\text{s}'' & \text{imposed heat flux}, \\
h(T_{\text{M}}-T_{\infty}) & \text{convective heat transfer}, \\
\varepsilon_{\text{R}} \sigma_{\text{R}}(T_{\text{M}}^4-T_{\text{R}}^4) & \text{radiative heat transfer},
\end{cases} \label{eq:thermal_BCs}
\end{equation}
with $h$ the convective heat transfer coefficient (W/m$^2$/K), $\varepsilon_{\text{R}}$ the emissivity of the surface, $\sigma_{\text{R}} = 5.67 \times 10^{-8}$~W/m$^2$/K$^4$ the Stefan-Boltzmann constant, $T_{\infty}$ the cooling fluid temperature~(K) and $T_{\text{R}}$ the radiative environment temperature (K).

The temperature field is discretized using nodal Lagrange elements and belongs to the space of square integrable functions in $\Omega_{\text{M,the}}$, such that $T_{\text{M}} \in \mathcal{H}^1(\Omega_{\text{M,the}})\times ]0,T_{\text{sim}} ]$. At the macroscopic scale, the thermal weak formulation reads: \\
From an initial solution at $t=0$, find $T_{\text{M}} \in \mathcal{H}^1(\Omega_{\text{M,the}})\times ]0,T_{\text{sim}} ]$ s.t.,
\begin{multline}
    (\rho c_p \partial_t T_{\text{M}}, T_{\text{M}}')_{\Omega_{\text{M,the}}} + (\bm{\kappa} \cdot \textbf{grad}~T_{\text{M}}, \textbf{grad}~T_{\text{M}}')_{\Omega_{\text{M,the}}} \\ + \langle \bar{f}(T_{\text{M}}), T_{\text{M}}' \rangle_{\Gamma_{\text{M,the}}} = (\bar{q}_{\text{s}}, T_{\text{M}}')_{\Omega_{\text{M,the}}}, \label{eq:macro_thermal_weak_form}
\end{multline}
$\forall T_{\text{M}}' \in \mathcal{H}^1(\Omega_{\text{M,the}})$, where $\langle\cdot,\cdot\rangle_{\Gamma}$ denotes the $L^2$ inner product over the domain boundary $\Gamma$. 

The time discretization of \eqref{eq:macro_thermal_weak_form} is also performed with the IE method. As thermal properties depend on temperature, an iterative fixed point scheme is performed at each time step of the macroscopic thermal simulation.

\subsection{Macroscopic magneto-thermal coupling}
\noindent The coupling between magnetic and thermal subproblems is achieved through the heat source term $\bar{q}_{\text{s}}$ in \eqref{eq:macro_thermal_weak_form} accounting for the losses. For the normal conducting parts of the cold mass, e.g. aluminum, the heat source term is given by Joule losses:
\begin{equation}
    \bar{q}_{\text{s}} = \sigma \lVert\bm{e}_{\text{M}}\rVert^2 = \sigma \lVert\partial_t \bm{a}_{\text{M}}\rVert^2. \label{eq:Joule_losses}
\end{equation}
The computation of the loss density $\bar{q}_{\text{s}}$ in the LTS coil is described in Section~\ref{sec:multi-scale-approach} and Section~\ref{sec:semi-analytical-approach}. \\
The $\sigma$ dependence on temperature is neglected in the normal conductors and the macroscopic magnetodynamic subproblem~\eqref{eq:macro_magnetic_weak_form} does not depend on the thermal one~\eqref{eq:macro_thermal_weak_form}. Thus, a one-way coupling scheme is considered, with the thermal subproblem being solved after the magnetodynamic one.

\section{Analytical Approximations for Losses in Multifilamentary LTS Conductors}
\label{sec:analytical-approximations}
\noindent Multifilamentary LTS conductors, as the wire-in-channel cable depicted in Fig.~\ref{fig:C400_geo}, are composed of a large number of SC filaments embedded inside a copper stabilizing matrix. In such wires, two types of losses are considered: filament hysteresis losses (AC losses) due to persistent currents in the SC filaments and inter-filament coupling losses \cite{Wilson}, due to current loops crossing the copper matrix between twisted filaments. Longitudinal Joule losses in the matrix are neglected.

\subsection{Filamentary Hysteresis Losses}
\begin{figure*}[!t]
    \centering
    \includegraphics{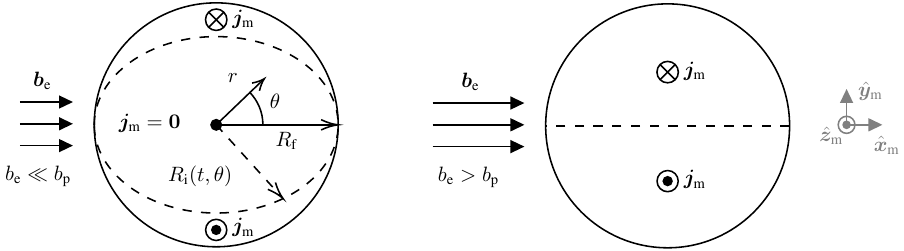}
    \caption{Expected distribution of the current density $\bm{j}_{\text{m}}$ in a superconducting cylinder of radius $R_\text{f}$ subjected to an uniform external increasing transverse flux density $\bm{b}_{\text{e}}$ of fixed direction, both in the weak (left) and in the full (right) penetration regimes, based on the CSM and adapted from \cite{Carr}. The inner radius of the current shell is denoted by $R_{\text{i}}(t,\theta)$. The norm of the external flux density $\bm{b}_{\text{e}}$ is denoted by $\lVert \bm{b}_{\text{e}} \rVert = b_{\text{e}}$, while $b_{\text{p}}$ represents the penetration flux density.}
    \label{fig:th_bg:PenetrationCurrent}
\end{figure*}

\noindent Hysteresis losses are due to persistent currents in the SC filaments, accounting for induced currents as well as for transport currents. Analytical approximations focus on the transverse field loss in a single SC filament of diameter $d_{\text{f}}$, denoted as $q_{\text{hys},1}$ (W/m$^3$). These approximations ignore the local twisting of the SC filaments and treat them as infinite cylinders at a local scale, which are reasonable assumptions for the considered magnets. Carr \cite{Carr} and Wilson \cite{Wilson} have described the hysteresis losses in a filament subjected to a cyclic transverse field based on Bean's Critical State Model (CSM) \cite{CSM}. Considering the magnet ramp-up, the analytical approximations are now extended to instantaneous losses in a ramping external field.

Under the assumptions of no transport current, an uniform external flux density $b_{\text{e}}$ and a constant critical current density~$j_{\text{c}}$, the transverse field loss in a fully penetrated filament is given by \cite{Carr}:
\begin{equation}
    q_{\text{hys},1} = \frac{2}{3\pi} d_{\text{f}} j_{\text{c}}  \dot{b}_{\text{e}}, \label{eq:hysteresis_loss_full_penetration}
\end{equation}
with $\dot{b}_{\text{e}} \triangleq \lVert \partial_t \bm{b}_{\text{e}} \rVert$. Equation~\eqref{eq:hysteresis_loss_full_penetration} is only valid above the penetration flux density ($b_{\text{e}} > b_{\text{p}}$), which can be approximated by $b_{\text{p}}~=~\mu_0~d_{\text{f}}~j_{\text{c}}~/\pi$ under the same assumptions. The corresponding current density configuration is shown on the right of Fig.~\ref{fig:th_bg:PenetrationCurrent}, with the $\cdot_{\text{m}}$ subscript referring to the fields at the mesoscopic scale. Based on the CSM, the electric field is given by $e_{\text{m},z} = - \dot{b}_{\text{e}} y_{\text{m}}$ in full penetration considering an external field applied along $\hat{\bm{x}}_{\text{m}}$.

The opposite asymptotic regime is the weak penetration regime ($b_{\text{e}} \ll b_{\text{p}}$). During the first magnetization, shielding currents are established on the outer shell of the filament to prevent the magnetic flux from entering the filament. The expected current density distribution \cite{Carr} is shown on the left of Fig.~\ref{fig:th_bg:PenetrationCurrent}. Assuming no transport current, a constant $j_{\text{c}}$ and a monotonic uniform field variation, the instantaneous hysteresis power loss in the weak penetration regime is given by:
\begin{equation}
    q_{\text{hys},1} = \frac{64}{3\pi d_{\text{f}} j_{\text{c}} \mu_0^2} b_{\text{e}}^2~\dot{b}_{\text{e}}, \label{eq:hysteresis_loss_weak_penetration}
\end{equation}
for which the full development is based on the CSM and can be found in Appendix B. 

The transition between the two regimes cannot be modelled analytically. As proposed in \cite{Carr} for cyclic fields, one can interpolate between the two regimes:
\begin{equation}
    q_{\text{hys},1} = \frac{1}{3\pi} \frac{2 d_{\text{f}} j_{\text{c}} b_{\text{e}}^2}{d_{\text{f}}^2 j_{\text{c}}^2 \mu_0^2 / 32 + b_{\text{e}}^2} \, \dot{b}_{\text{e}}, \label{eq:hysteresis_loss_full_range}
\end{equation}
which approximates the instantenous transverse hysteresis loss in the whole range of applied flux density $b_{\text{e}}$. Equation~\eqref{eq:hysteresis_loss_full_range} reduces to \eqref{eq:hysteresis_loss_full_penetration} for large $b_{\text{e}}$ and to \eqref{eq:hysteresis_loss_weak_penetration} for low~$b_{\text{e}}$.

The macroscopic loss density $q_{\text{hys}}$ (W/m$^3$) in the LTS coil is extrapolated from the filamentary hysteresis loss density \cite{WilsonBook}:
\begin{equation}
    q_{\text{hys}} = \lambda_{\text{SC}} \, q_{\text{hys},1}, \label{eq:macro_hysteresis_loss_full_range}
\end{equation}
with $\lambda_{\text{SC}}$ the SC filling ratio in the composite conductor.

\subsection{Inter-filament Coupling Losses}
\noindent Coupling losses are due to the magnetic flux variation over several filaments, which induces eddy currents in the copper matrix between filaments, referred to as coupling currents~\cite{PhysicalOriginCouplingCurrents}. Assuming that screening currents do not prevent field penetration into the composite material, Carr~\cite{Carr} derived an analytical expression for coupling losses~$q_{\text{c}}$ (W/m$^3$):
\begin{equation}
    q_{\text{c}} = \frac{\lambda_{\text{SC}}}{\lambda_{\text{st}}} \frac{1}{\rho_{\text{et}}} \left( \frac{p}{2\pi}\right)^2 \left\lVert\frac{d\bm{b}_{\text{e}}}{dt}\right\rVert^2, \label{eq:coupling_loss}
\end{equation}
with 
\begin{equation}
    \rho_{\text{et}} = \rho_{\text{Cu}} \frac{1+\lambda_{\text{st}}}{1-\lambda_{\text{st}}}, \label{eq:effective_transverse_resistivity}
\end{equation}
$\lambda_{\text{st}}$ the local SC filling ratio in the central strand, $\rho_{\text{et}}$ the effective transverse resistivity of the copper matrix ($\Omega$~m), $\rho_{\text{Cu}}$ the resistivity of copper ($\Omega$~m), and $p$ the filament twist pitch length (m). The effective transverse resistivity \eqref{eq:effective_transverse_resistivity} is determined \cite{Carr} assuming a large contact resistance at the matrix-filament interface, expected due to intermetallic layers formed during the fabrication process~\cite{WilsonBook}. The magnetoresistive effect of copper is taken into account by using empirical relationships proposed in \cite[pp. 8--23]{SimonMagnetoResistance}. 

\section{Description and Characterization of the Mesoscopic Model}
\label{sec:characterization-meso}
\noindent In practical applications characterized by slow ramp rates, inter-filament coupling losses are usually significantly lower than hysteresis losses~\cite[Fig. 16]{DularCATI}, as the former are proportional to the squared norm of the applied field variation. Given the extended ramp-up times of medical magnets over several hours, the focus is set on the accurate description of hysteresis losses. To this end, a FE model of a single filament is proposed to compute the AC losses at the mesoscopic scale inside the LTS coil, in order to provide a more accurate evaluation of $q_{\text{hys},1}$ than \eqref{eq:hysteresis_loss_full_range}. While a three-dimensional (3D) model of twisted filaments would offer greater accuracy of the AC losses at the mesoscopic scale, especially for faster ramp rates involving coupling losses, the proposed model is more simple and allows a fast computation of hysteresis losses at the filament scale. While this model is introduced to enable the cross-validation of the approximations presented in Section~\ref{sec:analytical-approximations} for estimating the filamentary hysteresis losses, it will also be used in Section~\ref{sec:multi-scale-approach} to implement efficiently the multi-scale approach.

\begin{figure*}[!t]
    \centering
    \includegraphics{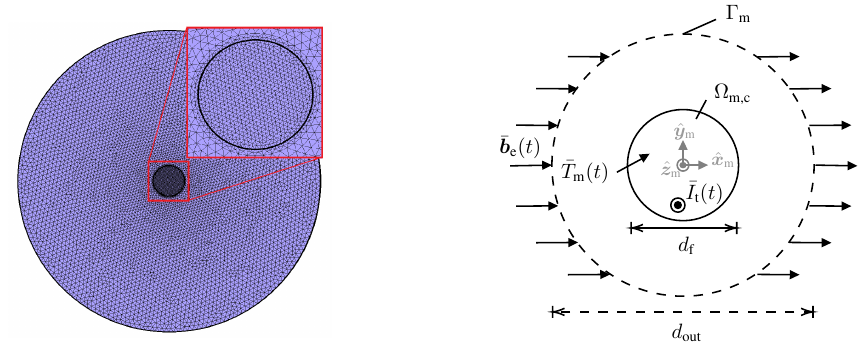}
    \caption{Mesoscopic submodel. Mesh discretization of the two-dimensional (2D) single filament (SF) model with a focus on the central superconducting filament (left) and the corresponding boundary conditions (right): uniform external flux density $\bar{\bm{b}}_{\text{e}}(t)$ applied to outer domain boundary $\Gamma_{\text{m}}$, applied transport current $\bar{I}_{\text{t}}(t)$ and uniform filament temperature $\bar{T}_{\text{m}}(t)$. The SC filament $\Omega_{\text{m,c}}$ of diameter $d_{\text{f}}$ is placed within a non-conducting domain $\Omega_{\text{m,c}}^{\text{C}}$ of fixed diameter $d_{\text{out}} = 10 \, d_{\text{f}}$. The mesh, corresponding to 5235 degrees of freedom (DOFs), has been refined and the convergence of AC loss predictions has been verified.}
    \label{fig:singleFilamentModel}
\end{figure*}

\subsection{Description of the Single Filament Model}
\noindent The single filament FE model, its mesh and the corresponding boundary conditions are represented in Fig.~\ref{fig:singleFilamentModel}. Neglecting the radius of curvature along the $\hat{\bm{z}}_{\text{m}}$-axis, the single filament is approximated as an infinite cylinder and its losses are evaluated with a 2D model. To assess the accuracy of the hysteresis losses, the outer domain is chosen as non-conducting. The power-law~(PL) \cite{powerLaw1, powerLaw2} is assumed to hold in the SC filament:
\begin{equation}
    \bm{e}_{\text{m}} = \rho_{\text{SC}}(\bm{j}_{\text{m}})~\bm{j}_{\text{m}} = \frac{e_{\text{c}}}{j_{\text{c}}} \left( \frac{\lVert \bm{j}_{\text{m}} \rVert}{j_{\text{c}}} \right)^{n-1} \bm{j}_{\text{m}}, \label{eq:power_law}
\end{equation}
with $e_{\text{c}}=10^{-4}$~V/m by convention. Even though the $n$ exponent is expected to decrease with $b$ and $T$ \cite{Godeke}, it is fixed to $n=50$ in this study, which is a reasonable value for Nb-Ti filaments in copper composites at $b < 5$~T \cite{GhoshPowerLaw}. Experimental observations in \cite{GhoshPowerLaw} suggest the power-law model offers a better representation of the electrical constitutive relationship in practical superconductors than the CSM.

The magnetic field $h$-$\phi$ formulation is used to solve the discretized electromagnetic problem at the scale of the filament, as it deals efficiently with the SC non-linearity~\cite{DularFEFormulationsSC}. The mesoscopic domain of interest is denoted by $\Omega_{\text{m}}$. The magnetic field $\bm{h}_{\text{m}}$ is approximated with edge basis functions in the filament $\Omega_{\text{m,c}}$. In the non-conducting domain~$\Omega_{\text{m,c}}^{\text{C}} = \Omega_{\text{m}} \setminus \Omega_{\text{m,c}}$, $\textbf{curl}~\bm{h}_{\text{m}} = \bm{0}$ and $\bm{h}_{\text{m}} = - \textbf{grad}~\phi_{\text{m}}$, such that $\bm{h}_{\text{m}}$ is approximated with gradients of nodal basis functions.
The mesoscopic magnetic weak formulation reads:
\\
From an initial solution at $t=0$, find $\bm{h}_{\text{m}} \in \mathcal{H}(\text{curl},\Omega_{\text{m}}) \times ]0,T_{\text{sim}}]$ s.t.,
\begin{multline}
    ( \mu_0\partial_t \bm{h}_{\text{m}}, \bm{h}_{\text{m}}')_{\Omega_{\text{m}}} +( \rho_{\text{SC}}~\textbf{curl}~ \bm{h}_{\text{m}}, \textbf{curl}~ \bm{h}_{\text{m}}')_{\Omega_{\text{m,c}}} \\ - \langle \bar{\bm{e}}_{\text{m}} \times \bm{n}, \bm{h}_{\text{m}}'\rangle_{\Gamma_{\text{m}}} = 0, \label{eq:meso_magnetic_weak_form}
\end{multline}
$\forall \bm{h}_{\text{m}}' \in \mathcal{H}(\text{curl},\Omega_{\text{m}})$, with $\textbf{curl}~\bm{h}_{\text{m}}' = \bm{0}$ in $\Omega_{\text{m,c}}^{\text{C}}$. 

The net transport current $\bar{I}_{\text{t}}(t)$ condition across the filament is applied strongly with cohomology basis functions as described in \cite{Pellikka}. The uniform external flux density $\bar{\bm{b}}_{\text{e}}(t)$ condition can be applied weakly by using the fact that the test function can be set to $\bm{h}_{\text{m}}' = -\textbf{grad}~\phi_{\text{m}}'$ on $\Gamma_{\text{m}}$. In this case, it has been shown in \cite[pp. 106--107]{PDularPhD} that the last term in~\eqref{eq:meso_magnetic_weak_form} can be replaced by:
\begin{align}
\langle \bar{\bm{e}}_{\text{m}} \times \bm{n}, \bm{h}_{\text{m}}'\rangle_{\Gamma_{\text{m}}} &= \langle \bm{n} \cdot (\textbf{curl}~ \bar{\bm{e}}_{\text{m}}), \phi_{\text{m}}'\rangle_{\Gamma_{\text{m}}} \nonumber \\
&= -\langle \bm{n} \cdot \partial_t \bar{\bm{b}}_{\text{e}}, \phi_{\text{m}}'\rangle_{\Gamma_{\text{m}}}, \label{eq:weak_form_meso_boundary_condition}
\end{align}
which amounts to enforce $\bm{n} \cdot \bar{\bm{b}}_{\text{e}}(t)$ as $\bar{\bm{b}}_{\text{e}}(t=0) = \bm{0}$. This does not interfere with the application of $\bar{I}_{\text{t}}(t)$ as transport current-induced field is azimuthal far from the SC filament.

The thermal problem is not solved at the mesoscopic scale and the temperature of the filament is assumed uniform and given by $\bar{T}_{\text{m}}(t)$. The filament temperature can be taken into account by considering the temperature dependence of the local critical current density $j_{\text{c}}(b_{\text{m}},\bar{T}_{\text{m}})$.

The time integration of \eqref{eq:meso_magnetic_weak_form} is performed with the IE method using an adaptive time stepping strategy as described in~\cite{JDularPhd}. At each mesoscopic time-step, the resulting non-linear system is solved with an iterative NR scheme. In cylindrical coordinates, the filamentary hysteresis power loss per unit volume $q_{\text{hys},1}$ is then evaluated as 
\begin{equation}
    q_{\text{hys},1} = \frac{4}{\pi d_{\text{f}}^2} \int_0^{d_{\text{f}}/2} \int_0^{2\pi} \bm{j}_{\text{m}} \cdot \bm{e}_{\text{m}}~ r~dr d\theta. \label{eq:powerLossEvaluation}
\end{equation}

\subsection{Cross-validation of the Hysteresis Loss Approximation} \label{subsec:cross-validation}
\noindent The accuracy of the CSM-based approximation \eqref{eq:hysteresis_loss_full_range} is evaluated through a comparison with the hysteresis loss density obtained numerically with the SF model. No temperature dependence is taken into account in this section. As a boundary condition, the uniform external flux density $\bar{b}_{\text{e}}$ is increased from 0 to $\bar{b}_{\text{e,max}} = 3$~T with a constant ramp rate~$\dot{b}_{\text{e}}$, which is varied in the following discussion. The direction of $\bar{b}_{\text{e}}$ is considered constant. The diameter of the filament is set to $d_{\text{f}} = 156$~$\mu$m. Different magnetic field dependences are considered for the critical current densities $j_{\text{c}}(b)$, and different transport current waveforms are considered for $\bar{I}_{\text{t}}(t)$.

\subsubsection{Constant critical current density without transport current}
A constant critical current density $j_{\text{c}} = 5\times 10^9$~A/m$^2$ is considered without any transport current, corresponding to the assumptions underlying the analytical approximation~\eqref{eq:hysteresis_loss_full_range}.

\begin{figure}[!t]
\centering
\includegraphics{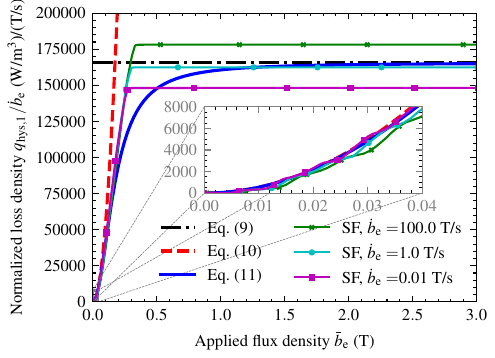}
\caption{Single filament hysteresis loss density $q_{\text{hys,1}}$ computed numerically with the SF model, for a monotonic applied field ramp-up at various ramp rates $\dot{b}_{\text{e}}$. The losses are normalized by the ramp rate and are compared to CSM-based analytical approximations: full penetration \eqref{eq:hysteresis_loss_full_penetration}, weak penetration \eqref{eq:hysteresis_loss_weak_penetration} and full-range interpolation \eqref{eq:hysteresis_loss_full_range}. Numerical parameters: $d_{\text{f}} = 156$~$\mu$m, $n=50$, cst. $j_{\text{c}}=5 \times 10^9$~A/m$^2$.}
\label{fig:cross_val_cst_jc}
\end{figure}

Numerical results for the filamentary hysteresis loss density are represented in Fig.~\ref{fig:cross_val_cst_jc} for various ramp rates $\dot{b}_{\text{e}}$ and are compared to the analytical approximations. It can be observed that the weak penetration approximation~\eqref{eq:hysteresis_loss_weak_penetration} is consistent with the results at low fields. The numerical oscillations observed can be attributed to the finite spatial discretization and the persisting currents penetrating the filament element by element. In the weak penetration regime, the losses seem to decrease when the ramp rate is increased, which is at odds with~\eqref{eq:hysteresis_loss_weak_penetration}.

While the full-range interpolation~\eqref{eq:hysteresis_loss_full_range} is accurate at large fields, it underestimates the losses in the intermediate field range. In particular, the numerical ratio between the hysteresis loss and the ramp rate becomes constant as soon as full penetration is reached (based on the CSM, $b_{\text{p}} = 0.312$~T). Also, the fully penetrated regime seems to be achieved at lower fields as the ramp rate is decreased. The normalized loss ratio $q_{\text{hys},1}/\dot{b}_{\text{e}}$ in full penetration does not correspond to the CSM-based approximation~\eqref{eq:hysteresis_loss_full_penetration} at high fields, as it increases with the ramp rate, while it is expected by~\eqref{eq:hysteresis_loss_full_penetration} to be rate independent.

The fully penetrated filamentary hysteresis loss can also be estimated analytically assuming $e_{\text{m},z} = - \dot{b}_{\text{e}} y_{\text{m}}$ holds in the PL model (for $\bar{\bm{b}}_{\text{e}}$ along $\hat{\bm{x}}_{\text{m}}$), which is equivalent to the constant filament magnetization assumption. A constant current density distribution would induce a constant self-field distribution within the filament, such that the local flux density variation would be given by $\partial_t \bm{b}_{\text{m}}=\partial_t\bm{b}_{\text{e}} = \dot{b}_{\text{e}}\hat{\bm{x}}_{\text{m}}$. In this context, \eqref{eq:powerLossEvaluation} can be evaluated ($y_{\text{m}} = r \sin{\theta}$) by taking advantage of the problem symmetry with respect to the $\hat{\bm{x}}_{\text{m}}$-axis:
\begin{align}
    q_{\text{hys},1} &= \frac{8}{\pi d_{\text{f}}^2} \int_0^{\pi} \int_0^{d_{\text{f}}/2} j_{\text{c}} \left(\frac{\dot{b}_{\text{e}} r \sin{\theta}}{e_{\text{c}}}\right)^{\frac{1}{n}} \dot{b}_{\text{e}}\,r \sin{\theta}~r~dr d\theta \nonumber \\
    &= \frac{8j_{\text{c}}\dot{b}_{\text{e}} (\dot{b}_{\text{e}} / e_{\text{c}})^{\frac{1}{n}}}{\pi d_{\text{f}}^2}  \int_0^{d_{\text{f}}/2} r^{\frac{2n+1}{n}}dr \int_{0}^{\pi} (\sin\theta)^{\frac{n+1}{n}}d\theta \nonumber \\
    &= \frac{8j_{\text{c}}\dot{b}_{\text{e}} (\dot{b}_{\text{e}} / e_{\text{c}})^{\frac{1}{n}}}{\pi d_{\text{f}}^2} \frac{(d_{\text{f}}/2)^{(3+1/n)}}{3+1/n}\int_{0}^{\pi} (\sin\theta)^{\frac{n+1}{n}}d\theta \nonumber \\ &= \frac{\int_{0}^{\pi}(\sin\theta)^{\frac{n+1}{n}}d\theta}{(3+1/n)\pi}\, j_{\text{c}}d_{\text{f}}\dot{b}_{\text{e}} \left( \frac{d_{\text{f}} \dot{b}_{\text{e}}}{2e_{\text{c}}}\right)^{\frac{1}{n}}. \label{eq:full_penetration_power_law}
\end{align}
As expected,~\eqref{eq:full_penetration_power_law} tends towards the CSM approximation~\eqref{eq:hysteresis_loss_full_penetration} for $n \rightarrow \infty$. The $\int_{0}^{\pi}(\sin\theta)^{(n+1)/n}d\theta/(3+1/n)$ factor in~\eqref{eq:full_penetration_power_law} is close to $2/3$ for high $n$ values. For $n=50$, it is equal to $0.6582$ which yields a $1.4$\% relative difference with respect to the CSM asymptotic regime ($2/3$).

\begin{figure}[!t]
\centering
\includegraphics{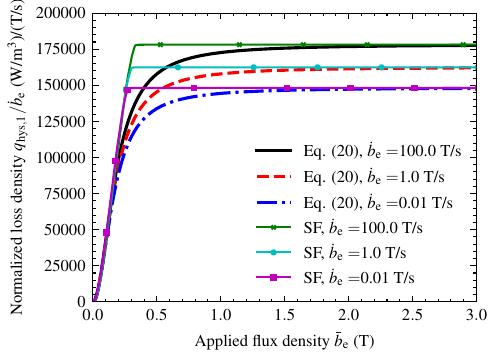}
\caption{Single filament hysteresis loss density $q_{\text{hys,1}}$ computed numerically with the SF model, for a monotonic applied field ramp-up at various ramp rates $\dot{b}_{\text{e}}$. The losses are normalized by the ramp rate and are compared to the analytical full-range approximation \eqref{eq:hysteresis_loss_full_range_update}. Numerical parameters: $d_{\text{f}} = 156$~$\mu$m, $n=50$, cst. $j_{\text{c}}=5 \times 10^9$~A/m$^2$.}
\label{fig:cross_val_cst_jc_update}
\end{figure}

To take into account this rate dependence, the full-range filamentary hysteresis loss approximation~\eqref{eq:hysteresis_loss_full_range} must be replaced~by
\begin{align}
    q_{\text{hys},1} &= \frac{2A}{3\pi} \frac{d_{\text{f}} j_{\text{c}} b_{\text{e}}^2}{d_{\text{f}}^2 j_{\text{c}}^2 \mu_0^2 A / 32 + b_{\text{e}}^2}\, \dot{b}_{\text{e}}, \label{eq:hysteresis_loss_full_range_update} \\
    \text{with}& \quad A = \frac{1}{2/3} \frac{\int_{0}^{\pi}(\sin\theta)^{\frac{n+1}{n}}d\theta}{3+1/n} \left( \frac{d_{\text{f}} \dot{b}_{\text{e}}}{2e_{\text{c}}}\right)^{\frac{1}{n}}. \nonumber
\end{align}
This final full-range approximation tends towards~\eqref{eq:full_penetration_power_law} at large fields, while still tending towards~\eqref{eq:hysteresis_loss_weak_penetration} at low fields. Predictions based on~\eqref{eq:hysteresis_loss_full_range_update} are compared to numerical results in Fig.~\ref{fig:cross_val_cst_jc_update}. In full penetration, theoretical predictions based on~\eqref{eq:full_penetration_power_law} are accurate as the maximal relative difference in fully penetrated losses is $0.085$\% with respect to the numerical results.

\subsubsection{Critical surface and transport current}
\noindent In practice, the critical current density varies with both the applied field and the temperature. Bottura's \cite{BotturaFit} scaling law $j_{\text{c}}(b,T)$ is now considered as it has been found to match experimental measurements. The corresponding fitting parameters are those obtained in~\cite{BotturaFit} with Spencer's data set~\cite{SpencerData}, Lubell's description of the critical flux density dependence on temperature \cite{LubellCriticalField} and $j_{\text{c}}(5~\text{T}, 4.2~\text{K}) = 2783$~A/mm$^2$.

\begin{figure}[!t]
    \centering
    \includegraphics{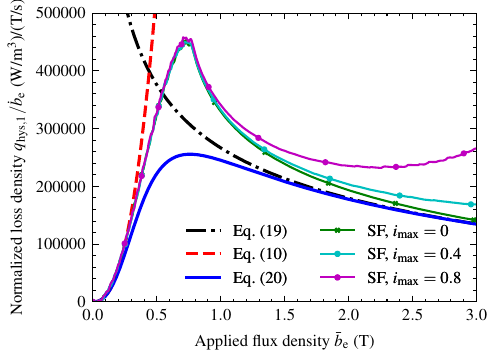}
    \caption{Single filament hysteresis loss density $q_{\text{hys,1}}$ computed numerically with the SF model, for a monotonic applied field ramp-up at ramp rate $\dot{b}_{\text{e}} = 0.01$~T/s and various linear applied transport current ratios $i_{\text{max}}$. The losses are normalized by the ramp rate and are compared to analytical approximations: PL-based full penetration \eqref{eq:full_penetration_power_law}, CSM-based weak penetration \eqref{eq:hysteresis_loss_weak_penetration} and full-range interpolation \eqref{eq:hysteresis_loss_full_range_update}. Numerical parameters: $d_{\text{f}} = 156$~$\mu$m, $n=50$, $j_{\text{c}}(b,\bar{T}_{\text{m}})$: Bottura's scaling law, with fixed $\bar{T}_{\text{m}}=4.2$~K.}
    \label{fig:cross_val_jcb}
\end{figure}

Numerical results obtained considering the local $j_{\text{c}}(b_{\text{m}},\bar{T}_{\text{m}})$ dependence are represented in Fig.~\ref{fig:cross_val_jcb}. They are compared to the analytical approximations~\eqref{eq:full_penetration_power_law} and~\eqref{eq:hysteresis_loss_full_range_update}, in which the $j_{\text{c}}$ factor is interpreted as $j_{\text{c}}(\bar{b}_{\text{e}},\bar{T}_{\text{m}})$. The temperature is kept constant at $\bar{T}_{\text{m}}=4.2$~K. Results are also represented for a linear increase in transport current $\bar{I}_{\text{t}}(t)$ over the ramp time fixed to $T_{\text{ramp}}=\bar{b}_{\text{e,max}}/\dot{b}_{\text{e}} = 300$~s. In this case, the transport current is imposed simultaneously with the applied field ramp and its slope is controlled through the parameter $ i_{\text{max}} \triangleq \bar{I}_{\text{t}}(T_{\text{ramp}}) / I_{\text{c}}(3~\text{T}, 4.2~\text{K})$.

In the absence of transport current, the filamentary hysteresis loss density is accurately described by the CSM-based weak penetration approximation~\eqref{eq:hysteresis_loss_weak_penetration} at low fields and by the PL-based full penetration approximation~\eqref{eq:full_penetration_power_law} at high fields. It highlights the validity of interpreting these equations with $j_{\text{c}} = j_{\text{c}}(\bar{b}_{\text{e}},\bar{T}_{\text{m}})$. In the fully-penetrated regime, the losses decrease as $\bar{b}_{\text{e}}$ is increased since $\partial j_{\text{c}}/\partial b_{\text{m}} < 0$ locally and the magnetization of the filament decreases. As a consequence, the loss density is maximal when full penetration is first reached. 

The full-range interpolation again underestimates the loss density in the intermediate applied field regime. In this regime, values of the normalized loss density obtained numerically are also larger than predictions with the fully penetrated approximation. This was not observed in the constant $j_{\text{c}}$ case and it may be explained by the local $j_{\text{c}}$ being larger in the center of the filament due to screening currents opposing the penetration of the flux density. As a result, the underestimation of the filamentary loss with the full-range approximation~\eqref{eq:hysteresis_loss_full_range_update} is more pronounced than in the simplified constant $j_{\text{c}}$ case.

With transport current, the situation is more complex to analyze. As highlighted in Fig.~\ref{fig:cross_val_jcb}, transport current leads to an increase in hysteresis losses. The full-range interpolation~\eqref{eq:hysteresis_loss_full_range_update} is not able to capture this effect as the transport current has been neglected during its derivation. The weak penetration approximation~\eqref{eq:hysteresis_loss_weak_penetration} is still consistent with the numerical results at low fields as the instantaneous transport current is negligible with respect to the critical current. In the fully penetrated regime, Wilson~\cite{WilsonBook} predicts a factor $(1+i^2)$ increase of the AC transverse field hysteresis loss when a DC transport current is applied, with the current ratio $i=\bar{I}_{\text{t}}/I_{\text{c}}$. This prediction is based on the CSM in the constant $j_{\text{c}}$ case. Carr~\cite{Carr} provides a similar, but more elaborated approximation. The $(1+i^2)$ factor is not retrieved numerically. It may be explained by the fact that neither the local flux density nor the local current density is uniform in the SC filament. Consequently, the impact of transport current cannot be predicted with a closed-form approximation.

In conclusion, the complexity of the multiple phenomena involved in the electromagnetic behaviour of the SC filament makes it difficult to provide a simple analytical approximation for the filamentary hysteresis loss density, particularly in the intermediate applied field regime.

\section{Multi-scale Approach}
\label{sec:multi-scale-approach}
\noindent As underlined above, analytical approximations for hysteresis losses may lack accuracy with respect to numerical models. The multi-scale approach presented in this section aims at providing an accurate prediction of hysteresis losses while avoiding an excessive computational cost.

\begin{figure}[!t]
\centering
\includegraphics{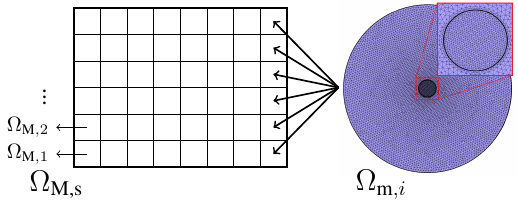}
\caption{Separation of scales in the context of the multi-scale approach: the cross-section of the macroscopic LTS coil $\Omega_{\text{M,s}}$ is subdivided into $N_z$ different zones $\Omega_{\text{M},1}$, ..., $\Omega_{\text{M},N_z}$. One mesoscopic model $\Omega_{\text{m},i}$ (here, the SF model introduced previously) is associated with each macroscopic zone $\Omega_{\text{M},i}$.}
\label{fig:separation_of_scales}
\end{figure}

\subsection{Principle of the Method}
\noindent The multi-scale approach relies on the separation between the macroscopic scale of the LTS coil and the mesoscopic scale of the SC filaments as depicted in Fig.~\ref{fig:separation_of_scales}. The main idea is to compute the hysteresis losses directly at the mesoscopic scale. To this end, the cross-section of the macroscopic LTS coil is subdivided into $N_z$ different zones $\Omega_{\text{M},1}$, ..., $\Omega_{\text{M},N_z}$. One generic mesoscopic model $\Omega_{\text{m},i}$ is associated with each zone for loss prediction.

While this approach is rather general, the SF model is chosen as the mesoscopic model in this study. This choice is motivated by the fact that inter-filament coupling losses are negligible with respect to hysteresis losses in the context of medical applications designed with magnets characterized by relatively slow ramp rates. This result is highlighted in Campbell's loss map~\cite[Fig. 10]{CampbellLossMap}, reproduced numerically with the CATI method~\cite[Fig. 15]{DularCATI}, showing that the filamentary hysteresis losses are dominant in the ultra-low frequency regime. In this regime, the SC filaments are uncoupled and the local current density distribution is similar in neighbouring filaments~\cite[Fig.~17(a)]{DularCATI}, as the screening of the applied magnetic field is weak. Consequently, working with one single filament model per zone allows the filamentary hysteresis losses to be computed with reasonable accuracy and to be extrapolated to the whole zone. By considering a zone subdivision that is coarser than the macroscopic mesh, the resolution is faster than with conventional multi-scale methods as the FE$^2$ method \cite{FE2Method} or the Homogeneous Multiscale Method \cite{HMM}, which associate one mesoscopic model with each macroscopic integration point.

\begin{figure}[!t]
    \centering
    \includegraphics{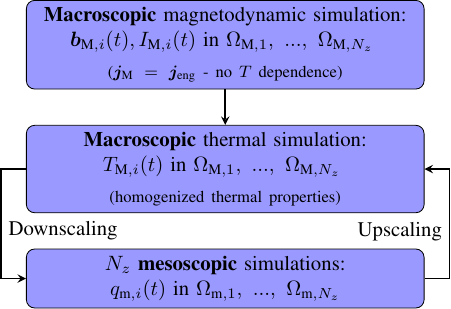}
    \caption{Global flowchart of the multi-scale procedure for computing the temperature distribution at the macroscopic scale of the LTS coil, while computing hysteresis losses at the mesoscopic scale of the SC filaments.}
    \label{fig:multi-scale_global_flowchart}
\end{figure}

The global flowchart of the multi-scale procedure is summarized in Fig.~\ref{fig:multi-scale_global_flowchart}. First, the macroscopic flux density distribution is obtained from a magnetodynamic simulation at the macroscopic scale (Section~\ref{sec:macro-mag}) with a fixed time step $\Delta t_{\text{M}}$. During post-processing, the average flux density $\bm{b}_{\text{M},i}(t_n)$ and the average transport current per filament $I_{\text{M},i}(t_n)$ are retrieved in each zone $\Omega_{\text{M},i}$ at each time step $t_n = n \Delta t_{\text{M}}$, with $n \in \{1,...,T_{\text{sim}}/\Delta t_{\text{M}}\}$. The volume average $x_{\text{M},i}$ of the generic macroscopic quantity $x_{\text{M}}$ in the zone $\Omega_{\text{M},i}$ is
\begin{equation}
    x_{\text{M},i} \triangleq \frac{1}{|\Omega_{\text{M},i}|} \int_{\Omega_{\text{M},i}} x_{\text{M}}~d\Omega_{\text{M},i}.
\end{equation}    
As the thermal dependence of the macroscopic magnetic properties is neglected, the magnetic simulation can be performed offline, prior to the multi-scale approach itself.

Next, the macroscopic thermal simulation (Section~\ref{sec:macro-the}) is performed with Joule losses \eqref{eq:Joule_losses} as the heat source in normal conducting parts $\Omega_{\text{M,c}}$. In the LTS coil $\Omega_{\text{M,s}}$, the volumetric heat source is defined as a piecewise constant function:
\begin{equation}
    \bar{q}_{\text{s}} = q_{\text{c}} + \bar{q}_{\text{M},i} \quad \text{with} \quad \bar{q}_{\text{M},i} = \lambda_{\text{SC}} q_{\text{m},i}, \label{eq:heatSource_coupled_multiScale}
\end{equation}
in which the coupling losses $q_{\text{c}}$ are computed with~\eqref{eq:coupling_loss} -- neglecting the magnetic response of the filaments, \eqref{eq:coupling_loss} is evaluated with $\bm{b}_{\text{e}} = \bm{b}_{\text{M}}$ -- and the mesoscopic hysteresis loss density $q_{\text{m},i}$ is the filamentary hysteresis loss density $q_{\text{hys},1}$~\eqref{eq:powerLossEvaluation} obtained from the mesoscopic simulation in the zone $\Omega_{\text{m},i}$.

While the mesoscopic simulation depends on the average $\bm{b}_{\text{M},i}(t)$ and $I_{\text{M},i}(t)$ sequences computed previously, it also depends on the $T_{\text{M},i}(t)$ sequence through the local critical current density $j_{\text{c}}(b_{\text{m}},\bar{T}_{\text{m},i}=T_{\text{M},i})$. Hence, both mesoscopic and macroscopic thermal simulations depend on each other. The thermal simulation is thus solved in parallel with the $N_z$ mesoscopic simulations. 

\begin{figure*}[!t]
\centering
\includegraphics{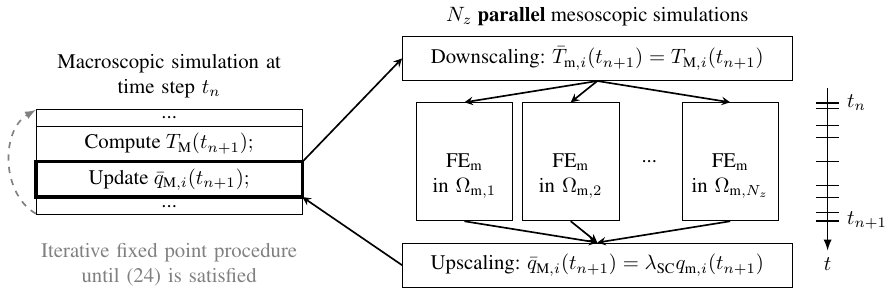}
\caption{Implementation of the iterative fixed point scheme in the multi-scale procedure. One macroscopic thermal resolution is performed before downscaling the temperature. In each mesoscopic zone $\Omega_{\text{m},i}$, the SF model simulation is restarted from the converged solution $\bm{h}_{\text{m},i}(t_n)$ of the previous time step. The $N_z$ mesoscopic simulations (denoted FE$_{\text{m}}$) are performed in parallel, until the next macroscopic time step $t_{n+1}$ is reached. The hysteresis loss is then upscaled. These fixed point iterations between scales are performed until the integrated hysteresis loss has converged. The time axis on the right of the figure is representative of the adaptive time stepping procedure at the mesoscopic scale. The time axis is not to scale as there can be several orders of magnitude between macroscopic and mesoscopic time steps.}
\label{fig:multi-scale_implementation}
\end{figure*}

A fixed point iteration between the scales is performed at each macroscopic time step $t_n$ until convergence of the integrated hysteresis loss is achieved.
At $t_n$, the goal is to predict the temperature distribution $T_{\text{M}}(t_{n+1})$, based on the heat source $\bar{q}_{\text{s}}(t_{n+1})$ (IE time-integration scheme) and on $\bar{q}_{\text{M},i}(t_{n+1})$ through~\eqref{eq:heatSource_coupled_multiScale}. At the beginning of each fixed point iterative procedure, the first iterate is set to $\bar{q}_{\text{M},i}(t_{n+1}) = \bar{q}_{\text{M},i}(t_{n})$. The implementation of one multi-scale fixed point iteration within GetDP is represented in Fig.~\ref{fig:multi-scale_implementation}. First, the updated macroscopic temperature distribution $T_{\text{M}}(t_{n+1})$ is computed, as well as its average $T_{\text{M},i}(t_{n+1})$ in each zone $\Omega_{\text{M},i}$. During downscaling, the average quantities $\bm{b}_{\text{M},i}(t_{n+1})$, $I_{\text{M},i}(t_{n+1})$ and $T_{\text{M},i}(t_{n+1})$ are passed to the mesoscopic simulations. At the mesoscopic scale, they correspond respectively (cf. Fig.~\ref{fig:singleFilamentModel}) to the applied field $\bar{\bm{b}}_{\text{e}}$, the transport current $\bar{I}_{\text{t}}$ and the average filament temperature $\bar{T}_{\text{m},i}$ at time $t_{n+1}$ in each zone $\Omega_{\text{m},i}$. The mesoscopic simulations are then restarted from the last converged mesoscopic magnetic field distribution $\bm{h}_{\text{m},i}(t_n)$ at the previous time step. The evolution with time of the average quantities $\bm{b}_{\text{M},i}$, $I_{\text{M},i}$ and $T_{\text{M},i}$ is assumed linear between $t_n$ and $t_{n+1}$. The flux density variation required in~\eqref{eq:weak_form_meso_boundary_condition} is approximated by $\dot{\bm{b}}_{\text{M},i} \approx (\bm{b}_{\text{M},i}(t_{n+1}) - \bm{b}_{\text{M},i}(t_n))/\Delta t_{\text{M}}$. The $N_z$ mesoscopic simulations are executed in parallel as they are independent. After the mesoscopic simulations computed the required filamentary hysteresis loss $q_{\text{m},i}(t_{n+1})$ in each zone $\Omega_{\text{m},i}$, it is then passed back to the macroscopic thermal simulation during upscaling.

At each macroscopic time step $t_n$, the iterative communication between the scales (down- and upscaling) takes place until convergence of the integrated hysteresis loss at macroscopic time step $t_{n+1}$:
\begin{equation}
    Q_{\text{hys}}(t_{n+1}) = \sum_{i=1}^{N_z} \bar{q}_{\text{M},i}(t_{n+1})\cdot |\Omega_{\text{M},i}|.
\end{equation} 
In practice, convergence is assumed when the relative difference between two successive fixed point iterations is lower than a fixed tolerance $\varepsilon_{Q}$:
\begin{equation}
    \left|\frac{\delta Q_{\text{hys}}(t_{n+1})}{Q_{\text{hys}}(t_{n+1})}\right| < \varepsilon_{Q}, \label{eq:convergence_criterion}
\end{equation}
with the practical value set to $\varepsilon_{Q} = 10^{-3}$ in this study. Once convergence is reached, one last macroscopic thermal simulation is performed to obtain the converged temperature distribution $T_{\text{M}}(t_{n+1})$.

With the proposed approach, the macroscopic thermal model can differ from the macroscopic magnetic model, with different meshes and different domains of interest (as long as $\Omega_{\text{M,the}} \subset \Omega_{\text{M,mag}}$), depending on corresponding boundary conditions. The multi-scale approach is thus flexible and can be adapted to various configurations.

Fig.~\ref{fig:multi-scale_implementation} highlights the finer time-stepping procedure required at the mesoscopic scale to accurately describe the flux penetration inside the SC filaments. Down- and upscaling, executing the mesoscopic (child) simulations in parallel, as well as storing the intermediate mesoscopic field distributions, are managed by Python scripts embedded within the macroscopic (parent) thermal GetDP solver.

\subsection{Range of Validity and Limitations}
\noindent Even though the proposed multi-scale method is quite general, its accuracy is limited by the assumptions made at the mesoscopic scale. In particular, the SF model is expected to provide accurate results for applications involving slow ramp rates, as inter-filament coupling losses are negligible with respect to hysteresis losses in the ultra-low frequency regime. The extension of the proposed approach to faster phenomena should consider a 3D model of twisted filaments at the mesoscopic scale as inter-filament coupling should be taken into account. The most rigorous option would be to implement a 3D model of a single wire-in-channel conductor at the mesoscopic scale, while considering the upscaling of the magnetic response of single conductors back to the macroscopic scale.

Apart from the choice of the mesoscopic model and from the macroscopic modelling assumptions themselves, several factors can lead to numerical errors:
\begin{itemize}
    \item Classical numerical errors due to the finite element discretization and the time-stepping scheme, along with the non-linear nature of the different problems at both scales. In particular, dealing with the SC non-linearity at the mesoscopic scale requires a significant computational effort, which makes the use of parallel computing essential.
    \item The coil cross-section subdivision into zones (cf. Fig.~\ref{fig:separation_of_scales}) may lead to errors in the hysteresis loss prediction. The choice of the number of zones $N_z$ is crucial and should be adapted to the coil geometry and the specific application.
    \item Due to smaller time steps at the mesoscopic scale, downscaled quantities must be interpolated in time. The choice of the interpolation method may have an impact on the convergence of the multi-scale procedure and sufficiently small macroscopic time steps should be used to mitigate this effect.
    
\end{itemize}

\section{Semi-analytical approach} \label{sec:semi-analytical-approach}
\noindent A semi-analytical approach is also proposed for computing the hysteresis losses in the LTS coil. It relies exclusively on the macroscopic magnetic and thermal formulations described in Section~\ref{sec:problem-description} and the analytical approximations respectively introduced and adapted in Section~\ref{sec:analytical-approximations} and Section~\ref{sec:characterization-meso}. In the LTS coil $\Omega_{\text{M,s}}$, the volumetric heat source is defined as:
\begin{equation}
    \bar{q}_{\text{s}} = q_{\text{c}} + \lambda_{\text{SC}} q_{\text{hys,1}}, \label{eq:heatSource_semiAnalytical}
\end{equation}
with $q_{\text{hys,1}}$ the filamentary hysteresis loss density computed with the full-range analytical approximation~\eqref{eq:hysteresis_loss_full_range_update}, in which the critical current density is evaluated as $j_{\text{c}}(b_{\text{M}},T_{\text{M}})$. As the PL model is expected to be more accurate than the CSM for practical superconductors,~\eqref{eq:hysteresis_loss_full_range_update} is preferred over~\eqref{eq:hysteresis_loss_full_range}. The coupling losses $q_{\text{c}}$ are still computed with~\eqref{eq:coupling_loss}. Focusing on the low frequency range, the magnetic response of the filaments is again neglected, such that the external field is assumed equal to the macroscopic field $\bm{b}_{\text{e}}=\bm{b}_{\text{M}}$ in the evaluation of the analytical approximations.

The limitations of the semi-analytical approach arise from the assumptions inherent in the analytical approximation~\eqref{eq:hysteresis_loss_full_range_update}, as it neglects both the transport current and the $j_{\text{c}}(b,T)$ dependence. Moreover, the full-range interpolation does not provide accurate results in the intermediate applied field regime as highlighted in Section~\ref{sec:characterization-meso}. As a consequence, the semi-analytical approach is expected to be less accurate than the multi-scale approach. Nevertheless, it is easier to implement and it should be less computationally expensive.

\begin{figure*}[!t]
    \centering
    \includegraphics{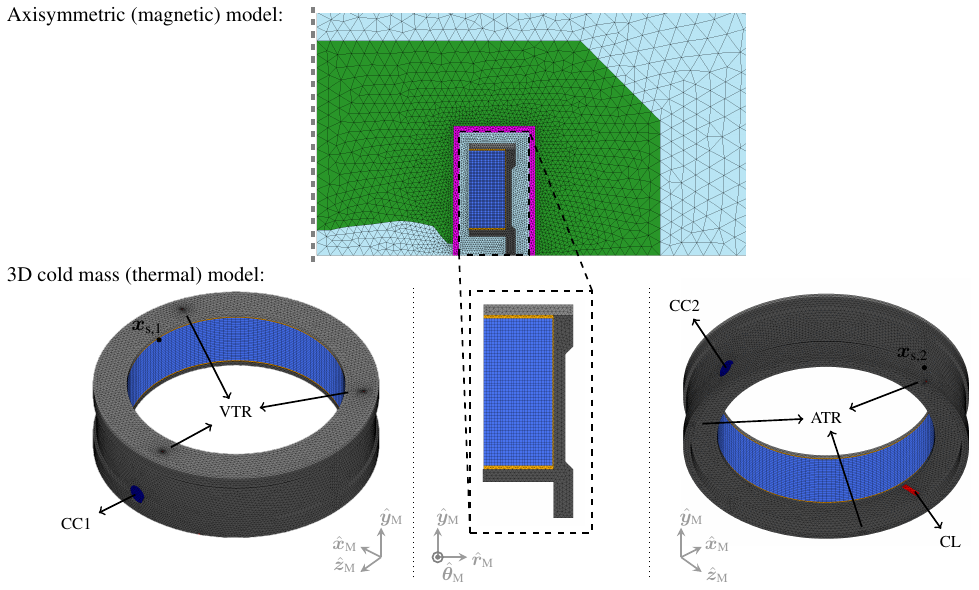}
    \caption{Top: Macroscopic magnetic (axisymmetric) model of the S2C2 cyclotron, along with its mesh (6003 DOFs). The cylindrical coordinate system is denoted ($\hat{\bm{r}}_{\text{M}}$,$\hat{\bm{y}}_{\text{M}}$,$\hat{\bm{\theta}}_{\text{M}}$). The cold mass is constituted of the LTS coil (navy blue), the aluminum coil former (dark grey), the stainless steel coil former flanges (light grey) and the epoxy resin insulating junction (orange). It is placed within an iron ferromagnetic yoke (green) supported by the cryostat wall (magenta). Remaining regions are filled with air/vacuum (light blue). For scale, the yoke diameter is $d_{\text{y}} = 2.5$~m. \\ Bottom: Macroscopic thermal 3D model of the cold mass $\Omega_{\text{M,the}}$ (top view on the left and bottom view on the right), along with its mesh (16720 DOFs). The cartesian coordinate system is denoted ($\hat{\bm{x}}_{\text{M}}$,$\hat{\bm{y}}_{\text{M}}$,$\hat{\bm{z}}_{\text{M}}$).
    The dark blue circular surfaces represent the two cryocooler (CC) surfaces. The red surfaces represent the imposed incoming heat flux surfaces: electrical current leads (CL), vertical tie rods (VTR) and axial tie rods (ATR). The positions of the two thermal sensors are denoted $\bm{x}_{\text{s,1}}$ and $\bm{x}_{\text{s,2}}$.
    }
    \label{fig:2D_s2C2}
\end{figure*}

\section{Application and Results}
\label{sec:results}
\noindent The proposed modelling approaches are applied to the magneto-thermal simulation of the LTS magnet of the S2C2 synchrocyclotron \cite{S2C2} developed by IBA. The S2C2 superconducting synchrocyclotron is used to accelerate ions as part of a compact proton therapy treatment system. It relies on the conduction cooling mechanism and the dry magnet technology. The focus of this section is set on the AC losses generated inside the coil during its ramp-up from zero to nominal current. Numerical predictions, obtained both with the multi-scale approach (MSA) and with the semi-analytical approach (SAA), are compared to experimental data.

\subsection{Geometry and Material Properties}
\begin{figure}[!t]
    \centering
    \includegraphics{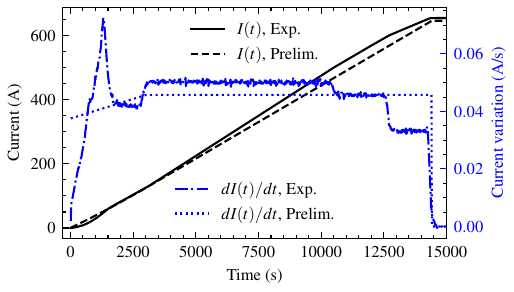}
    \caption{Current $I(t)$ passing through the LTS coil during the ramp-up procedure, along with its derivative with respect to time. Both the preliminary study~(Prelim.) current ramp-up and the experimentally (Exp.) measured current are represented. The experimental current has been smoothed out with a Savitzky-Golay filter \cite{SavitzkyGolayFilter} of window length $15$ and polynomial order $2$ to avoid excessive numerical variations linked to the current sensor resolution.}
    \label{fig:currentData}
\end{figure}

\noindent The complete 3D geometry of the S2C2 cyclotron can be found in~\cite[Fig. 3]{S2C2}. As the present study focuses on the cold mass of the cyclotron, the external components, the extraction systems and the cryocoolers are left out of the macroscopic magnetic model depicted in Fig.~\ref{fig:2D_s2C2}. The median plane symmetry is used to model the top half of the cyclotron. The LTS coil is made of 3128 wire-in-channel conductor turns arranged in series. The nominal current reaches $650$~A in 4 hours ($T_{\text{up}}\!=\!14400$~s). The two different current ramp-up procedures considered in this study are represented in Fig.~\ref{fig:currentData}, together with their time derivatives. From Section~\ref{subsec:numerical_results} to Section~\ref{subsec:iteration_study}, the preliminary ramp-up procedure $I(t)$ is used as input to discuss numerical properties of the multi-scale approach. In Section~\ref{subsec:exp_comparison}, the experimental current ramp-up (measured through a shunt resistor placed in series with the coil) is imposed in the numerical model to ensure a consistent comparison to experimental data. As observed, the current ramp-up is not linear as the current variation is smaller in the first phase due to the large initial inductance of the magnet.

The axisymmetric macroscopic magnetic model considers the experimental $h(b)$ curve of the ferromagnetic (iron) yoke. Apart from the yoke, $\mu=\mu_0$ is assumed everywhere. Eddy currents are considered in the coil former ($\sigma\!=\!78\!\times\!10^6$~S/m), in the flanges ($\sigma\!=\!2.06\!\times\!10^6$~S/m) as well as in the yoke ($\sigma\!=\!8\!\times\!10^6$~S/m). The Nb-Ti conductor parameters are gathered in Table~\ref{tab:coil_parameters}, with Bottura's $j_{\text{c}}(b,T)$ scaling law considered numerically as described in Section~\ref{sec:characterization-meso}.

\begin{table}
\begin{center}
\caption{Parameters of the Nb-Ti conductor in the S2C2 coil.}
\label{tab:coil_parameters}
\vspace*{-0.5em}
\begin{tabular}{ccccc}
\toprule
$\lambda_{\text{SC}}$ & $\lambda_{\text{st}}$ & $p$ & $d_{\text{f}}$ & $j_{\text{c}}(5~\text{T}, 4.2~\text{K})$\\
(-) & (-) & (m) & ($\mu$m) & (A/mm$^2$)  \\
\midrule
$0.148$ & $0.61$ & $0.1$ & $156$ & $2783$ \\
\bottomrule
\end{tabular}
\end{center}
\vspace*{-1.5em}
\end{table}

As the coil relies on conduction cooling (via four cryocoolers), a 3D thermal model of the cold mass is built as represented in Fig.~\ref{fig:2D_s2C2}. Experimentally, two Cernox\texttrademark~temperature sensors~\cite{Cernox} are placed on each sub-coil (one sub-coil is above the median plane, the other one below). The first sensor is located at the hottest point of the LTS coil and along the vertical midplane between the two cooling surfaces. The second sensor is located on the same plane, on the external edge of the coil former. Their respective locations are denoted by $\bm{x}_{\text{s},1}$ and $\bm{x}_{\text{s},2}$. 

Each sub-coil is cooled down by two Sumitomo RDK 415D cryocoolers at 50 Hz (cooling power map provided by the manufacturer \cite{CoolingCapacityMap-RDK415D}). The behaviour of the cryocoolers is complex as their cooling power depends on both first stage (radiation shield and other components at intermediate temperature) and second stage (cold mass) conditions. In particular, the second stage cooling power simultaneously depends on the second stage temperature and on the heat load at the first stage.
The simplifying assumption is made to model each cryocooler second stage contribution as a \textit{convective-like} boundary condition (BC). That is, each cryocooler cooling power $\bar{F}$ (W) is independent of the first stage conditions and is modelled as a linear function of the cold end temperature~$T_{\text{M}}$:
\begin{equation}
    \bar{F} = \tilde{h} \cdot (T_{\text{M}} - T_{\text{cryo}}) = h S_{\text{cryo}} (T_{\text{M}} - T_{\text{cryo}}), \label{eq:cooling_power}
\end{equation}
with $S_{\text{cryo}}$ the cold end surface (m$^2$), $h$ the convection coefficient to be used in the BC~\eqref{eq:thermal_BCs} and the parameters given by $\tilde{h}=1.45$~W/K and $T_{\text{cryo}} = 3.15$~K. 

Different heat sources, corresponding to the conduction through the tie rods and through the current leads (CL), together with the heat dissipation within the CL, are also modelled as fixed imposed heat fluxes on localized contact surfaces. On the remaining surface of the cold mass, an incoming radiative BC is imposed from the radiation shield. The median plane corresponds to an adiabatic surface.

The thermal material properties of the homogenized LTS coil are detailed in Appendix A. The thermal conductivity and specific heat (along with their thermal dependences) of the coil former ($\rho = 2700$ kg/m$^3$), the flanges ($\rho = 8000$ kg/m$^3$), and the insulating junction ($\rho = 1800$ kg/m$^3$) are taken from the NIST database~\cite{NIST_database} under the entries \textit{Aluminum 6061-T6}, \textit{Stainless Steel 304}, and \textit{Fiberglass Epoxy G-10 (normal direction)} respectively.

In the following, the initial cold mass temperature distribution is set as the steady-state temperature distribution obtained by solving~\eqref{eq:macro_thermal_weak_form} with $\bar{q}_{\text{s}} = 0$.

\subsection{Numerical results} \label{subsec:numerical_results}
\begin{figure}[!t]
\centering
\includegraphics[width=0.45\textwidth]{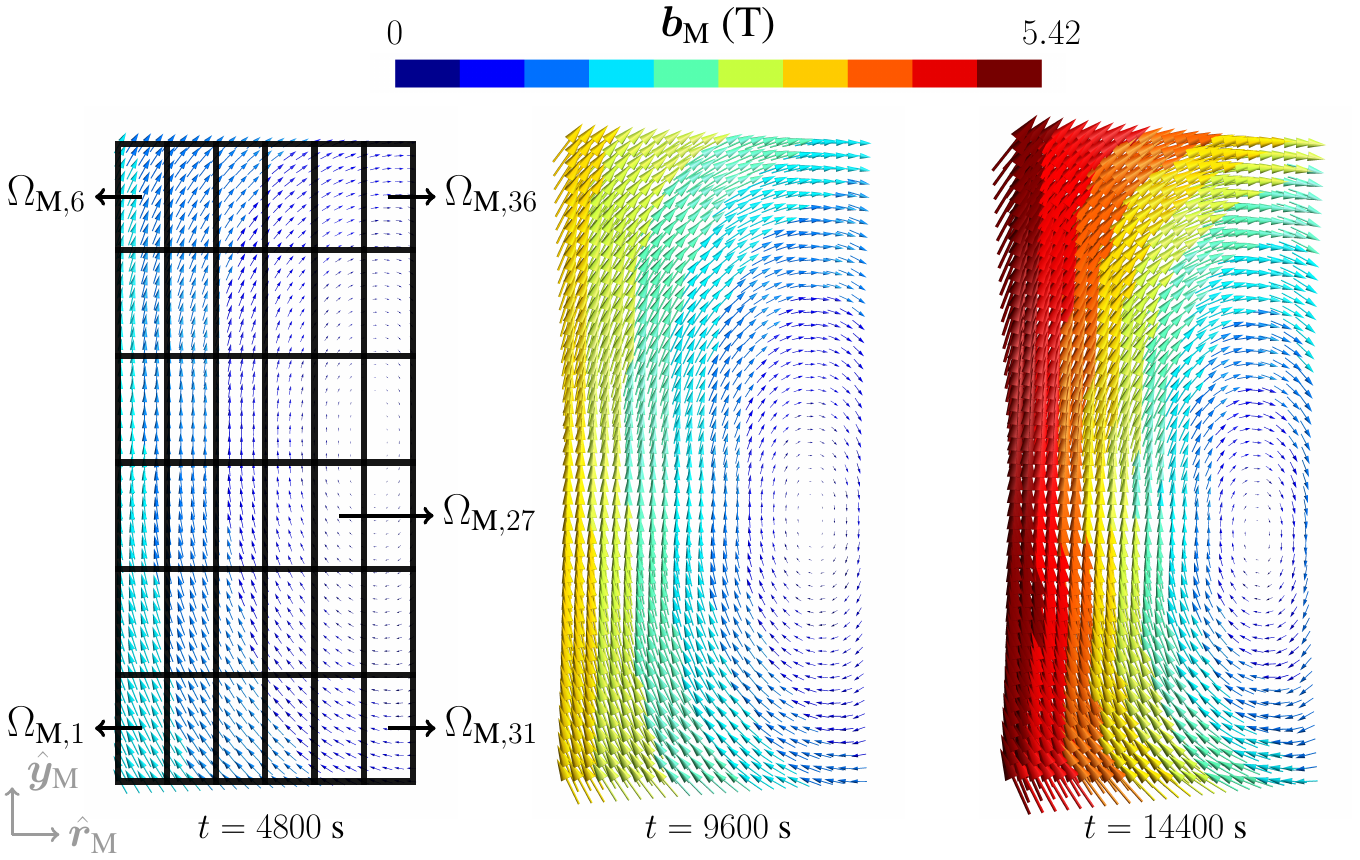}
\caption{Evolution of the macroscopic flux density distribution $\bm{b}_{\text{M}}$ within the LTS coil cross-section during the ramp-up procedure. Results obtained with the axisymmetric model. The left figure also shows the zone subdivision in the case of the MSA with $N_z=36$ zones.}
\label{fig:bMap}
\end{figure}

\noindent The macroscopic flux density distribution is obtained from the preliminary magnetodynamic simulation. Its evolution during the ramp-up procedure is represented in Fig.~\ref{fig:bMap}, with the focus set on the LTS coil. As may be observed, the part that is closest to the center of the cyclotron exhibits the largest field variations as the final flux density is larger along the inner diameter of the coil than along its outer diameter. Notably, the zero field region is crossing zone $\Omega_{\text{M},27}$ as it is moving towards the median plane during the ramp-up procedure. This is a direct consequence of the non-linearity of the macroscopic magnetic problem. The computed flux density distribution is then used as an input for the MSA.

\begin{figure}[!t]
\centering
\includegraphics{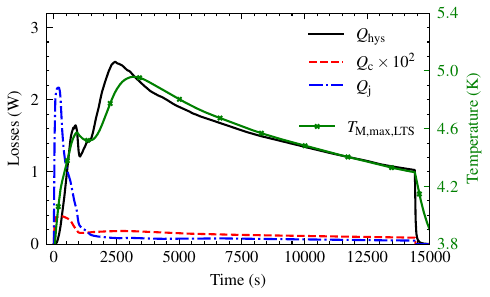}
\caption{Different loss contributions during the S2C2 ramp-up procedure: integrated hysteresis loss $Q_{\text{hys}}$, coupling loss $Q_{\text{c}}$ generated in the LTS coil, and Joule loss $Q_{\text{j}}$ generated in the coil former. The maximal temperature evolution $T_{\text{M,max,LTS}}$ in the LTS coil is also represented. Results are obtained with the MSA ($N_z=36$ zones).}
\label{fig:lossContributions}
\end{figure}

The evolution of the different loss contributions during the ramp-up procedure, computed with the MSA, is shown in Fig.~\ref{fig:lossContributions}. As highlighted, the coupling losses are negligible with respect to the hysteresis losses, a result which satisfies the fundamental assumption of the MSA. Joule losses in the coil former are predominant in the very first part of the ramp-up ($t\le 600$~s), before decreasing rapidly as the yoke saturates. Then, hysteresis losses in the LTS coil represent the main heat source during most of the ramp-up procedure, as Fig.~\ref{fig:lossContributions} highlights the strong correspondance between the hysteresis losses and the maximal temperature rise within the coil. Again, this shows the necessity of predicting these losses with great accuracy. 

\begin{figure}[!t]
\centering
\includegraphics{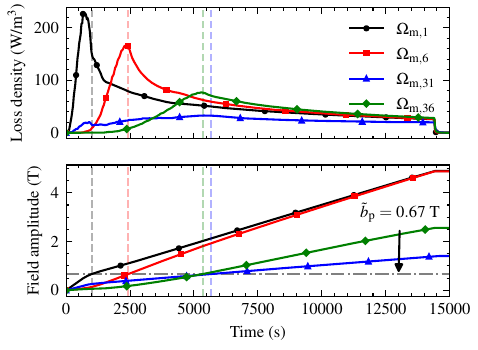}
\caption{Filamentary hysteresis loss density $q_{\text{m},i}$ (top) and corresponding average macroscopic flux density amplitude $b_{\text{M},i}$ (bottom) during the ramp-up procedure, for different zone indices $i \in \{1,6,31,36 \}$. Results are obtained with the MSA ($N_z=36$).}
\label{fig:hysLossExplanation}
\end{figure}

The two peaks in the hysteresis losses, around $t=1000$~s and $t=2500$~s, may be explained by the contribution of different zones in the coil. As can be deduced from Fig.~\ref{fig:cross_val_jcb} discussed previously, the filamentary loss is maximal when the applied flux density reaches $b_{\text{e}} = 0.67$~T~$\triangleq\tilde{b}_{\text{p}}$. Even though the approximate penetration flux density $\tilde{b}_{\text{p}}$ has been obtained assuming $T=4.2$~K, without transport current for an external field of constant direction and constant variation rate, similar conclusions can be drawn when considering the realistic conditions as shown in Fig.~\ref{fig:hysLossExplanation}. When the average flux density amplitude reaches the value of $\tilde{b}_{\text{p}}$ within a given zone, the corresponding filamentary hysteresis loss density is almost maximal. As a consequence, the two peaks in the integrated hysteresis losses $Q_{\text{hys}}$ correspond to the crossing of the penetration flux density in different zones, namely in the bottom and top zones along the inner diameter of the coil, respectively. These zones exhibit the largest field variations, thus contributing the most to $Q_{\text{hys}}$ as the filamentary hysteresis losses are in first approximation proportional to the field variation, see~\eqref{eq:hysteresis_loss_full_range_update}.

\subsection{Convergence of the Multi-scale Approach as a Function of the Number of Zones} \label{subsec:zone_convergence}
\begin{figure}[!t]
\centering
\includegraphics{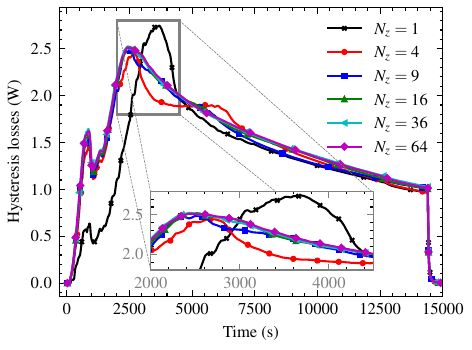}
\caption{Integrated hysteresis loss $Q_{\text{hys}}$ in the LTS coil during the ramp-up procedure with the MSA for various number of zones~$N_z$.}
\label{fig:zoneConv}
\end{figure}

\noindent As mentioned, the convergence of the results with the number of zones $N_z$ must be checked when using the MSA. In the present study, the zone subdivision is performed with a regular grid as shown in the left part of Fig.~\ref{fig:bMap} for $N_z=36$. The variation of the integrated hysteresis loss prediction with respect to $N_z$ is represented in Fig.~\ref{fig:zoneConv}. As expected, the integrated hysteresis loss $Q_{\text{hys}}$ converges as the number of zones is increased. For $N_z=1$, the MSA provides a poor hysteresis loss estimation, as the $Q_{\text{hys}}(t)$ evolution is significantly different from what is observed for larger $N_z$. This single-zone model smoothes out local $\bm{b}_{\text{M}}$ variations and therefore it does not capture the physics of the problem. In this configuration, the drop in losses around $t=1000$~s corresponds to a decrease in $\dot{b}_{\text{M},1}$ due to the saturation of the yoke. Results obtained with $N_z=4$ and $N_z=9$ also present spurious oscillations which highlight the necessity of further refining the zone subdivision.

Starting from $N_z=16$, the relative difference between successive zone refinements is becoming negligible as the different curves are nearly superimposed. Still, one can observe a slight increase in instantaneous hysteresis losses as $N_z$ is increased. Again, considering more zones allows for a better resolution of the local fields.

Note that the local oscillation in $Q_{\text{hys}}(t)$ observed in Fig.~\ref{fig:zoneConv} may be explained by different factors, such as the finite spatial discretization of the SF model at the mesoscopic scale and the discontinuous applied field variation induced by the linear interpolation of the macroscopic quantities over time.

\begin{table}
    \begin{center}
    \caption{Integrated hysteresis loss dissipation in the LTS coil predicted during the ramp-up procedure with the MSA for various number of zones $N_z$.}
    \label{tab:coil_zone_conv}
    \vspace*{-0.5em}
    \begin{tabular}{cc|cccc}
    \toprule
    $N_z$ & (-) & $1$ & $4$ & $9$ & $16$ \\
    \hline
    $E_{\text{hys}}$ & (J) & $19737$ & $21380$ & $21549$ & $22015$ \\
    \midrule
    $N_z$ & (-) & $25$ & $36$ & $64$ & $81$\\
    \hline
    $E_{\text{hys}}$ & (J) & $22266$ & $22253$ & $22252$ & $22298$ \\
    \bottomrule
    \end{tabular}
    \end{center}
    \vspace*{-1.5em}
\end{table}

The convergence of the results is confirmed by the numerical values of the integrated hysteresis loss $E_{\text{hys}}~=~\int_0^{T_{\text{up}}} Q_{\text{hys}}\, dt$ presented in Table~\ref{tab:coil_zone_conv}, as the relative variation of $E_{\text{hys}}$ does not exceed $0.2$\% for $N_z \in [25,81]$. It shows that one can consider a number of zones that is smaller than the number of integration points in the macroscopic mesh, thus accelerating the computation with respect to conventional multi-scale methods. The following results are presented with $N_z=36$ zones, as this choice ensures satisfactory accuracy while keeping a reasonable number of processing units required for the parallel computation.

\subsection{Number of Required Fixed Point Iterations} \label{subsec:iteration_study}
\begin{figure}[!t]
\centering
\includegraphics{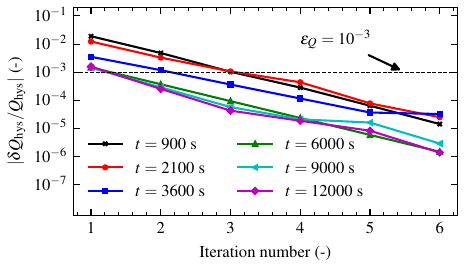}
\caption{Relative variation of the integrated hysteresis losses $|\delta Q_{\text{hys}} / Q_{\text{hys}}|$ between successive MSA iterations at various instants of the ramp-up procedure. Results are obtained with $N_z=36$. Results are shown with up to 6 fixed point iterations at each macroscopic time step, regardless of the stopping criterion~\eqref{eq:convergence_criterion}.}
\label{fig:MSiterConv}
\end{figure}

\noindent As explained in Section~\ref{sec:multi-scale-approach}, the MSA relies on fixed point iterations between scales to deal with the intrinsic non-linear behaviour of hysteresis losses. The relative variation of hysteresis losses between successive iterations is represented in Fig.~\ref{fig:MSiterConv}. As may be observed, the convergence is monotonic as the relative variation decreases with the number of iterations. The number of required iterations to reach the $\varepsilon_Q$ tolerance is larger in the first part of the ramp-up, in which the temperature variation is more important as inferred from Fig.~\ref{fig:lossContributions}. In the present study, the convergence is reached after $2$ to $4$ iterations. The strong convergence properties of the MSA can be attributed to the stabilizing behaviour of hysteresis losses. Indeed, they can be considered as proportional to $j_{\text{c}}$ in first approximation, cf. \eqref{eq:hysteresis_loss_full_range_update}, while $j_{\text{c}}$ is a decreasing function of the temperature. Hence, the hysteresis losses are expected to decrease with the temperature, all other parameters being kept constant. This result is specific to the normal operation of the LTS coil and can not be generalized to other configurations such as the thermal runaway in case of a quench event.

Moreover, Fig.~\ref{fig:MSiterConv} shows that one iteration is sufficient to reach a relative error of less than $2$\% on the hysteresis losses. In the final part of the ramp-up procedure, the single iteration relative error is even lower than $0.2$\%. This result is of particular interest if one aims for a faster computation of the hysteresis losses with the MSA.

\subsection{Comparison to Experimental Data} \label{subsec:exp_comparison}
\begin{figure*}[!t]
    \centering
    \includegraphics{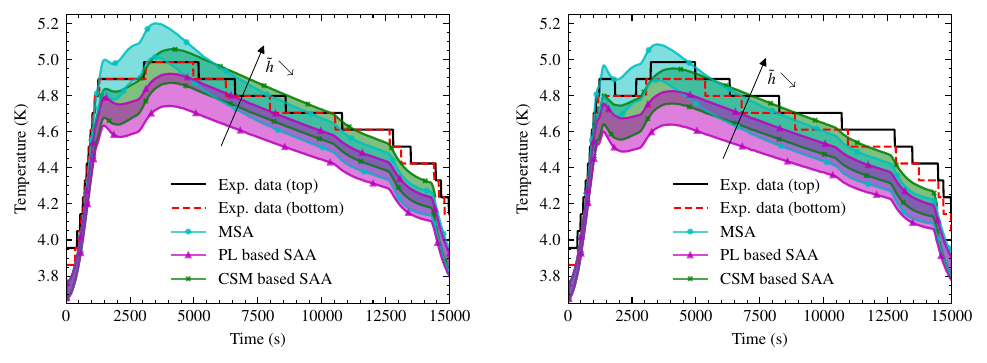}
    \caption{Temperature evaluated at the first sensor location $T_{\text{M}}(\bm{x}_{\text{s},1})$ (left) and at the second sensor location $T_{\text{M}}(\bm{x}_{\text{s},2})$ (right) during the ramp-up procedure. The experimental (Exp.) data is represented for both top and bottom sub-coils (respectively above and below the median plane, sensor resolution: $\delta T_{\text{s}} = 0.094$~K, $\delta t_{\text{s}}=15$~s), together with numerical results: MSA with $N_z=36$ zones, PL based SAA (based on~\eqref{eq:hysteresis_loss_full_range_update}), and CSM based SAA (based on~\eqref{eq:hysteresis_loss_full_range}). The cryocooler convective coefficient is varied between $\tilde{h}=0.97$~W/K and $\tilde{h}=1.45$~W/K.}
    \label{fig:exp_results}
\end{figure*}

\noindent Fig.~\ref{fig:exp_results} presents the comparison between the experimental data and the numerical results obtained both with the MSA and the SAA. As mentioned, the current imposed through the LTS coil is measured experimentally (cf. Fig.~\ref{fig:currentData}) and used as an input in the numerical simulations. As the cryocooler convective coefficient $\tilde{h}$ represents a significant source of uncertainty, the numerical results are showed for its expected value $\tilde{h}=1.45$~W/K (in optimal conditions) as well as for a $33$\% reduction of this value, i.e. $\tilde{h}=0.97$~W/K. The latter value is chosen as a realistic value (from the cryocooler cooling power map) for the lower bound of the expected range of $\tilde{h}$. Indeed, the first stage conditions may vary significantly during the ramp-up procedure and may influence the second stage cooling power at $4.2$~K. In particular, the Joule dissipation in the current leads is expected to significantly increase the first stage heat load, in which case the second stage cooling power is expected to be reduced. Moreover, the value of $\tilde{h}=0.97$~W/K leads to an initial temperature error lower than $0.1$~K between the numerical steady-state and the experimental temperature distribution, corresponding to the sensor resolution.

As may be observed, there is a slight asymmetry between the top and bottom experimental values. It highlights the uncertainty linked to the experimental measurements. The numerical results obtained with the MSA are in good agreement with the experimental data as the maximal temperature error is lower than $0.2$~K. Notably, the temperature rise predicted with the MSA is larger than the predictions obtained with the power-law based SAA. This result is expected as the mesoscopic SF model takes into account more physical phenomena than the analytical approximation~\eqref{eq:hysteresis_loss_full_range_update}. 

Moreover, both the MSA and the PL based SAA allow to numerically reproduce the shape of the experimental curves. The first temperature rise and the corresponding temperature peak observed at $\bm{x}_{\text{s},2}$ around $t=1500$~s has been predicted by the numerical models. The second temperature peak around $t=3500$~s is also reproduced numerically. However, the temperature rise is underestimated in the second-part of the ramp-up procedure as the numerical simulations predict a larger cooling rate than observed in the experiments. This discrepancy may be attributed, among other factors, to the non-linear behaviour of the cryocoolers as their cooling capacity may decrease during ramp-up.

Some tests have been performed with a CSM based SAA, which relies on~\eqref{eq:hysteresis_loss_full_range} instead of~\eqref{eq:hysteresis_loss_full_range_update} for the filamentary hysteresis loss within the LTS coil. The results obtained with the CSM based SAA are in better agreement with the experimental data than the PL based SAA, as the predicted temperature rise is around $0.1$~K larger. Also, the thermal behaviour in the second part of the ramp-up is reproduced more accurately. This result highlights the impact of the underlying SC constitutive model, even though the power-law is expected to better match experimental observations on single SC samples~\cite{GhoshPowerLaw} compared to the CSM. Results shown in Fig.~\ref{fig:exp_results} raise the question of the validity of the power-law model at the SC filament scale. The electric field within the SC filaments is very low during the second part of the ramp-up procedure involving large time scales. Different constitutive laws, e.g. the percolation model (mainly used for describing HTS), could be better suited to handle these slower phenomena involving relaxation \cite{ShenGrilliOverviewH, SiroisComparisonModels}.

Overall, the results presented in Fig.~\ref{fig:exp_results} allow to validate the implementation of both the MSA and the SAA proposed in this study.

\subsection{Discussion}
\begin{figure}[!t]
    \centering
    \includegraphics{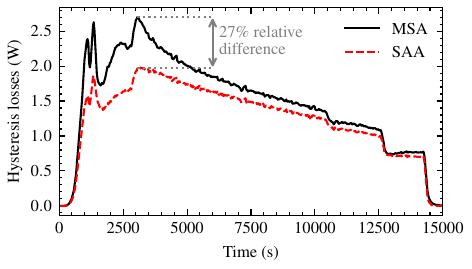}
    \caption{Integrated hysteresis loss $Q_{\text{hys}}$ in the LTS coil during the ramp-up procedure. Results are obtained with the MSA ($N_z=36$) and the PL based SAA, both with $\tilde{h} = 1.45$~W/K.}
    \label{fig:loss_MS_vs_SA}
\end{figure}

\noindent The hysteresis losses predicted with both the MSA and the power-law based SAA, corresponding to the temperature rise shown in Fig.~\ref{fig:exp_results}, are compared in Fig.~\ref{fig:loss_MS_vs_SA}. The fast loss oscillations may be induced by the corresponding $dI/dt$ oscillations in Fig.~\ref{fig:currentData}. Again, the correspondance between the losses in Fig.~\ref{fig:loss_MS_vs_SA} and the temperature rise in Fig.~\ref{fig:exp_results} is striking. The loss peaks around $t=1000$~s and $t=2500$~s can be explained by the crossing of the penetration flux density as discussed in Section~\ref{subsec:numerical_results}. The peak around $t=1400$~s is a consequence of the $dI/dt$ peak in the current ramp-up procedure observed in Fig.~\ref{fig:currentData}. Also, each significant $dI/dt$ variation in Fig.~\ref{fig:currentData} yields a corresponding hysteresis loss variation in Fig.~\ref{fig:loss_MS_vs_SA}, as these are in first approximation proportional to the field variation.

\begin{figure}[!t]
    \centering
    \includegraphics{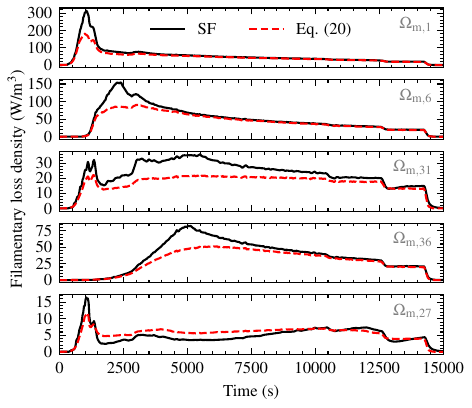}
    \caption{Filamentary hysteresis loss density $q_{\text{m},i}$ computed with the mesoscopic SF model during the ramp-up procedure, for different zone indices $i \in \{1,6,31,36,27 \}$. The prediction based on the analytical approximation~\eqref{eq:hysteresis_loss_full_range_update} is shown for comparison. Results are obtained with the MSA ($N_z=36$) and $\tilde{h} = 1.45$~W/K.}
    \label{fig:hysLossZones_MSA_vs_SAA}
\end{figure}

As observed in Fig.~\ref{fig:loss_MS_vs_SA}, the SAA underestimates the hysteresis losses with respect to the MSA, particularly in the first part of the ramp-up procedure in which losses are maximal. This observation can be inferred from the discussion in Section~\ref{subsec:cross-validation} as the analytical approximation~\eqref{eq:hysteresis_loss_full_range_update} was found to underestimate the losses in the intermediate field range with respect to the SF model. This behaviour is retrieved locally in most zones of the LTS coil as shown in Fig.~\ref{fig:hysLossZones_MSA_vs_SAA}, in which the analytical approximation again underestimates the losses. The largest losses are again occurring along the coil inner diameter. The interpretation of the results in zone $\Omega_{\text{m},27}$ is more complex as the local SC filament never reaches full penetration.

\begin{figure}[!t]
    \centering
    \includegraphics{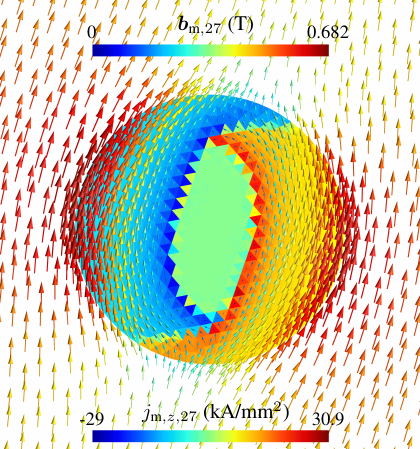}
    \caption{Mesoscopic flux density $\bm{b}_{\text{m},27}$ and current density out-of-plane component $j_{\text{m},z,27} = \bm{j}_{\text{m},27} \cdot \hat{\bm{z}}_{\text{m}}$ distributions in $\Omega_{\text{m},27}$ associated with the macroscopic zone $\Omega_{\text{M},27}$ shown in Fig.~\ref{fig:bMap}. Results correspond to the filament state after the current ramp-up procedure at $t=T_{\text{up}}$.}
    \label{fig:rotatingFields}
\end{figure}

Apart from the correct evaluation of the losses in the intermediate field range, the MSA can handle more complex phenomena and history as it goes beyond the assumptions of the analytical approximation. This is highlighted in Fig.~\ref{fig:rotatingFields} as the SF model can handle a rotating macroscopic magnetic field. Indeed, the remarkable current density distribution in the SC filament associated with zone $\Omega_{\text{M},27}$ is due to the macroscopic zero field region crossing the zone during ramp-up (cf. Fig.~\ref{fig:bMap}), such that the field is locally rotating. A similar distribution of field lines has been obtained numerically~\cite{Prigozhin} in a cylinder of square cross-section, subjected to a rotating magnetic field in the weak penetration regime. This particular behaviour is not captured by~\eqref{eq:hysteresis_loss_full_range_update} and it explains the discrepancy in the results shown in Fig.~\ref{fig:hysLossZones_MSA_vs_SAA} for $\Omega_{\text{m},27}$.

\begin{figure}[!t]
    \centering
    \includegraphics{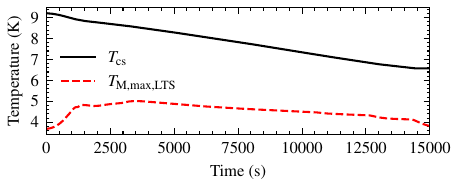}
    \caption{Minimal current-sharing temperature $T_{\text{cs}}$ and maximal temperature $T_{\text{M,max,LTS}}$ in the LTS coil during the ramp-up procedure. Results are obtained with the MSA ($N_z=36$) and $\tilde{h} = 1.45$~W/K.}
    \label{fig:Tcs}
\end{figure}

Still, the SAA yields satisfactory results when looking at the integrated hysteresis loss in Fig.~\ref{fig:loss_MS_vs_SA}. This may be attributed to the negligible impact of transport current for this particular application, as the maximal transport current ratio $i=\bar{I}_{\text{t}} / \max_{\bm{x} \in \Omega_{\text{M,s}}} I_{\text{c}}(b_{\text{M}},T_{\text{M}}) = 0.149$, occurring at the end of the ramp-up, is small. This highlights the large magnet operation margin with respect to the critical surface. Indeed, the corresponding temperature margin represented in Fig.~\ref{fig:Tcs} is important as well. The current-sharing temperature $T_{\text{cs}}$ is here defined as $I_{\text{c}}(b_{\text{M}},T_{\text{cs}}) \triangleq \bar{I}_{\text{t}}$.

\begin{figure}[!t]
    \centering
    \includegraphics{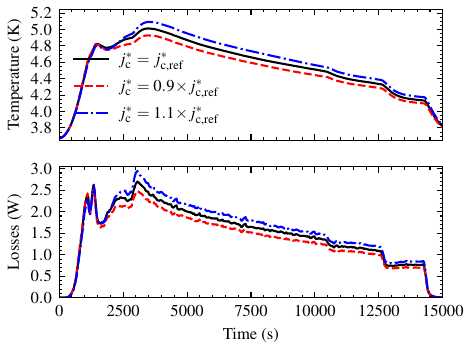}
    \caption{Temperature evaluated at the first sensor location $T_{\text{M}}(\bm{x}_{\text{s},1})$ (top) and corresponding integrated hysteresis loss $Q_{\text{hys}}$ in the LTS coil (bottom) during the ramp-up procedure. The critical current density scaling is varied between $j_{\text{c}}^* \triangleq j_{\text{c}}(5~\text{T}, 4.2~\text{K}) = 0.9\!\times\!j_{\text{c,ref}}^*$ and $j_{\text{c}}^* = 1.1\!\times\!j_{\text{c,ref}}^*$, with $j_{\text{c,ref}}^* = 2783$~A/mm$^2$. Results are obtained with the MSA ($N_z=36$) and $\tilde{h} = 1.45$~W/K.}
    \label{fig:jc_impact}
\end{figure}

While the uncertainty associated with the cryocooler convective coefficient and its influence on numerical results have been addressed in Section~\ref{subsec:exp_comparison}, there remain other uncertain parameters that could affect the results due to the highly non-linear nature of the problem. As an illustrative example, the accuracy of the critical current density $j_{\text{c}}(b,T)$ law, along with the corresponding values provided by the manufacturer, may have a significant impact as shown in Fig.~\ref{fig:jc_impact}. A $10$\% variation of the critical current density leads to a maximal temperature variation of $0.1$~K at the first sensor location. This is a direct consequence of the strong dependence of the hysteresis losses on $j_{\text{c}}$. As represented in Fig.~\ref{fig:jc_impact}, hysteresis losses also experience a $10$\% variation in the second part of the ramp-up, once most zones of the LTS coil are associated with fully penetrated filaments. In such cases, the filamentary loss is approximately proportional to $j_{\text{c}}$ (cf.~\eqref{eq:full_penetration_power_law}). In contrast, in the weak field regime, losses are less sensitive to $j_{\text{c}}$. The results in Fig.~\ref{fig:jc_impact} highlight the crucial role played by the accurate determination of SC material properties in the hysteresis loss prediction.

Even though the MSA yields a more accurate prediction of hysteresis losses, the SAA offers a significant advantage in terms of computing performance as highlighted in Table~\ref{tab:final_CPU_time}. Considering the stopping criterion~\eqref{eq:convergence_criterion} for the MSA, the computing time is significantly increased with respect to the SAA. However, using a single fixed point iteration between scales already ensures a relative error of less than 2\%, as pointed out in Section~\ref{subsec:iteration_study}. In this configuration, the MSA computing time can be greatly reduced while ensuring satisfactory accuracy. The computational effort required for the MSA is large, yet it remains much more advantageous than performing a full 3D simulation of the LTS coil down to the SC filament scale. The latter would require a meshing of the LTS coil with a resolution of the order of a tenth of the SC filament diameter, which is not feasible. In terms of memory usage, the MSA is again more demanding than the SAA, the parallel computation of several mesoscopic simulations requiring a significant amount of storage resources.

\begin{table}
    \begin{center}
    \caption{Computing (Wall) time required for a ramp-up simulation, together with the associated number of cores for the parallel computation and the corresponding memory usage (maximum resident set size). The MSA ($N_z = 36$) is performed both with $\varepsilon_Q = 10^{-3}$ and with a single fixed point (FP) iteration between scales. Time is reported for the different simulations: macroscopic magnetodynamic (M,mag), macroscopic thermal, both within SAA (M,the) and within MSA including mesoscopic~(M,the~+~m). Simulations were run on 32 core AMD Epyc Rome 7542 CPUs at 2.9 GHz.}
    \label{tab:final_CPU_time}
    \vspace*{-0.5em}
    \begin{tabular}{c|c|cc}
    \toprule
     & \multirow{2}{*}{SAA} & \multicolumn{2}{c}{MSA ($N_z = 36$)} \\ & & $\varepsilon_Q = 10^{-3}$ & single FP it. \\
    \midrule
    M,mag & $21$~min & \multicolumn{2}{c}{$26$~min} \\
    M,the & $3$~h $51$~min & / & / \\
    M,the + m & / & $27$~h $57$~min & $14$~h $48$~min \\ 
    \midrule
    Wall Time & $4$~h $12$~min & $28$~h $23$~min & $15$~h $14$~min \\
    \midrule
    Number of cores & 1 & 40 & 40 \\
    \midrule
    Memory usage & 392 Mb & 12.15 Gb & 12.07 Gb \\ 
    \bottomrule
    \end{tabular}
    \end{center}
    \vspace*{-1.5em}
\end{table}

\section{Conclusion}
\label{sec:conclusion}
\noindent In this work, a multi-scale approach (MSA) and a semi-analytical approach (SAA) were proposed to predict AC losses in LTS coils. They were implemented within magneto-thermal FE simulations using the GetDP open-source software. The application of these methods to the ramp-up procedure of the S2C2 LTS coil validated their implementation, as the numerical results were found to be in good agreement with experimental data.

Analytical approximations for the filamentary hysteresis losses were adapted to deal with ramping field boundary conditions. The cross-validation with a numerical single filament (SF) model showed that analytical approximations are accurate in asymptotic regimes, yet they underestimate the losses in the intermediate field range. Moreover, the critical current density dependence on the flux density and the effect of the transport current were highlighted as crucial factors than cannot be modelled with analytical approximations.

The MSA, relying on the SF model for loss prediction rather than analytical approximations, is expected to yield more accurate predictions compared to the SAA. Indeed, the SAA was found to underestimate losses when compared to the MSA, particularly in the intermediate field range. Consequently, numerical results obtained using the MSA exhibited better agreement with the experimental data than those obtained with the SAA. However, the SAA remains a valuable tool for fast computations, offering satisfactory accuracy in capturing the experimental curve shapes. While the computational cost associated with the MSA is higher, it still represents a more efficient alternative compared to conducting full 3D simulations of the LTS coil down to the scale of superconducting (SC) filaments. The MSA effectively accounts for mesoscopic phenomena within the SC filaments inside the LTS coil while maintaining a reasonable computational load.

Finally, it should be noted that the observed good agreement with the experimental data is a direct consequence of the negligible contribution of coupling losses with respect to hysteresis losses. An extension of the proposed MSA should consider replacing the mesoscopic SF model with a comprehensive conductor model, which would enable the description of faster phenomena involving inter-filament coupling losses.

\newpage

{\appendices
\section*{Appendix A: Homogenized Thermal Properties at Macroscopic Scale}
\begin{figure}[!t]
\centering
\includegraphics{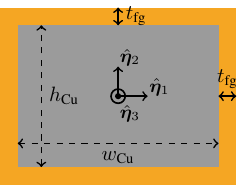}
\caption{Simplified cross-section of the wire-in-channel conductor (cf. Fig.~\ref{fig:C400_geo}), neglecting Nb-Ti. The height and the width of the copper channel are respectively denoted by $h_{\text{Cu}}$ and $w_{\text{Cu}}$, and the thickness of the fiberglass epoxy insulation by $t_{\text{fg}}$. The local coordinate system $\hat{\bm{\eta}}_1$-$\hat{\bm{\eta}}_2$-$\hat{\bm{\eta}}_3$ is also represented.}
\label{fig:app1}
\end{figure}
\noindent Neglecting the contribution of Nb-Ti, the homogenized thermal properties of the LTS coil are derived from the simplified geometry of a single wire-in-channel conductor as depicted in Fig.~\ref{fig:app1}. The (normalized, as Nb-Ti is neglected) filling factors of copper and fiberglass epoxy are respectively denoted by $\lambda_{\text{Cu}}$ and $\lambda_{\text{fg}}$. The homogenized density and the (mass-based averaged) homogenized specific heat capacity are respectively given by:
\begin{align}
    \rho_{\text{hom}} &= \lambda_{\text{Cu}} \rho_{\text{Cu}} + \lambda_{\text{fg}} \rho_{\text{fg}}, \\
    c_{p,\text{hom}} &= \frac{\lambda_{\text{Cu}} \rho_{\text{Cu}} c_{p,\text{Cu}} + \lambda_{\text{fg}} \rho_{\text{fg}} c_{p,\text{fg}}}{\rho_{\text{hom}}}.
\end{align}
The derivation of the homogenized thermal conductivity relies on the concept of thermal resistance \cite{HeatBook}. The distinction is made between the normal (radial and vertical) and warp (azimutal) fiberglass epoxy thermal conductivity ($\kappa_{\text{fg,n}}$ and $\kappa_{\text{fg,w}}$). Neglecting the radius of curvature of the conductor, the homogenized thermal conductivity along $\hat{\bm{\eta}}_1$ and $\hat{\bm{\eta}}_2$ results from the composite association of thermal resistances:
\begin{multline}
    \kappa_{1,\text{hom}} = \frac{w_{\text{Cu}} + 2 t_{\text{fg}}}{h_{\text{Cu}} + 2 t_{\text{fg}}} \left[ 2 \frac{t_{\text{fg}}}{\kappa_{\text{fg,n}} (h_{\text{Cu}} + 2 t_{\text{fg}}) } + \right.\\ \left. \left( \frac{\kappa_{\text{Cu}}h_{\text{Cu}}}{w_{\text{Cu}}} + 2 \frac{\kappa_{\text{fg,n}}t_{\text{fg}}}{w_{\text{Cu}}} \right)^{-1} \right]^{-1},
\end{multline}
\vspace*{-1em}
\begin{multline}
    \kappa_{2,\text{hom}} = \frac{h_{\text{Cu}} + 2 t_{\text{fg}}}{w_{\text{Cu}} + 2 t_{\text{fg}}} \left[ 2 \frac{t_{\text{fg}}}{\kappa_{\text{fg,n}} (w_{\text{Cu}} + 2 t_{\text{fg}}) } + \right.\\ \left. \left( \frac{\kappa_{\text{Cu}}w_{\text{Cu}}}{h_{\text{Cu}}} + 2 \frac{\kappa_{\text{fg,n}}t_{\text{fg}}}{h_{\text{Cu}}} \right)^{-1} \right]^{-1}.
\end{multline}
Along $\hat{\bm{\eta}}_3$, it results from the parallel association of thermal resistances:
\begin{equation}
\kappa_{3,\text{hom}} = \frac{\kappa_{\text{Cu}}h_{\text{Cu}}w_{\text{Cu}} + 2 \kappa_{\text{fg,w}} t_{\text{fg}}(h_{\text{Cu}} + w_{\text{Cu}} + 2 t_{\text{fg}})}{(h_{\text{Cu}} + 2 t_{\text{fg}})(w_{\text{Cu}} + 2 t_{\text{fg}})}.
\end{equation}
In the cylindrical coordinate system ($\hat{\bm{r}}_{\text{M}}$,$\hat{\bm{y}}_{\text{M}}$,$\hat{\bm{\theta}}_{\text{M}}$), the anisotropic homogenized thermal conductivity tensor of the LTS coil is then given by:
\begin{equation}
    \bm{\kappa}_{\text{hom}} = \begin{pmatrix} \kappa_{r} & 0 & 0 \\ 0 & \kappa_{y} & 0 \\ 0 & 0 & \kappa_{\theta} \end{pmatrix} = \begin{pmatrix} \kappa_{1,\text{hom}} & 0 & 0 \\ 0 & \kappa_{2,\text{hom}} & 0 \\ 0 & 0 & \kappa_{3,\text{hom}} \end{pmatrix}.
\end{equation}
The thermal conductivity and specific heat (along with their thermal dependences) of copper ($\rho = 8960$ kg/m$^3$) and fiberglass epoxy ($\rho = 1800$ kg/m$^3$) are retrieved from the NIST database~\cite{NIST_database} under the entries \textit{Copper OFHC (RRR100)} and \textit{Fiberglass Epoxy G-10} respectively.

\section*{Appendix B: Ramping Transverse Field Hysteresis Loss in Weak Penetration}
\noindent The derivation of the weak penetration hysteresis loss in a transverse ramping field is adapted from the work of Carr~\cite{Carr}. Assuming a virgin initial state and a monotonic field ramp-up, the current density distribution is shown in Fig.~\ref{fig:th_bg:PenetrationCurrent}, with $R_{\text{i}}(t,\theta)$ the moving boundary between $j=0$ and $j=j_{\text{c}}$ regions. For conciseness, the implicit $\cdot_{\text{m}}$ subscript is omitted in the following. In cylindrical coordinates $(r,\theta,z) \in \mathbb{R}^+ \times [ 0; 2\pi[ \times \mathbb{R}$, we have $h_z = e_{\theta} = e_r = 0$ by symmetry. The filamentary loss density~\eqref{eq:powerLossEvaluation} reduces to:
\begin{equation}
    q_{\text{hys},1} = 2 \cdot \frac{4}{\pi d_{\text{f}}^2} \int_0^{\pi} \int_0^{d_{\text{f}}/2} e_z \cdot j_z\,r\,dr d\theta, \label{eq:app2:powerLoss1}
\end{equation} 
such that only the $\theta \in [0;\pi[$ interval is considered next. Ampère's law and the radial component of Faraday's law respectively simplify to:
\begin{align}
    \frac{1}{r} \partial_r(r h_{\theta}) - \frac{1}{r} \partial_{\theta} h_r &= j_z, \label{eq:app2:ampere} \\
    \partial_r e_z = \mu_0 \dot{h}_{\theta}. \label{eq:app2:faraday}
\end{align}
Near the moving boundary $R_{\text{i}}$, still inside the filament, the current density is known from the CSM as:
\begin{equation}
    j_z = - j_{\text{c}}\,H(r-R_{\text{i}}) =
    \begin{cases} -j_{\text{c}} & \text{if } r > R_{\text{i}} \\ 0 & \text{if } r < R_{\text{i}},
    \end{cases}\label{eq:app2:j_z}
\end{equation}
with $H(\cdot)$ the Heaviside step function, whose derivative is the Dirac Delta $\delta(\cdot)$. Deriving~\eqref{eq:app2:ampere} with respect to time and introducing $R_{\text{i}} < r^* < d_{\text{f}}/2$:
\begin{align}
    \frac{1}{r} \partial_r(r \dot{h}_{\theta}) &- \frac{1}{r} \partial_{\theta} \dot{h}_r = \dot{j}_z = j_{\text{c}}\,\dot{R}_{\text{i}}\,\delta(r-R_{\text{i}}) \\
    \Rightarrow \int_{R_{\text{i}}^-}^{r^*} \dot{j}_z\,r\,dr &= j_{\text{c}}\,\dot{R}_{\text{i}}\,R_{\text{i}} \nonumber \\ &= r^*\,\dot{h}_{\theta}  - \left(R_{\text{i}}\,\dot{h}_{\theta}\right)^- - \int_{R_{\text{i}}^-}^{r^*} \partial_{\theta} \dot{h}_r\,dr \nonumber \\
    &= r^*\,\dot{h}_{\theta} \label{eq:app2:h_theta_closed_form1} \\
    \Rightarrow \dot{h}_{\theta} &= j_{\text{c}}\,\dot{R}_{\text{i}}\frac{R_{\text{i}}}{r^*} \approx j_{\text{c}}\,\dot{R}_{\text{i}} \label{eq:app2:h_theta_closed_form2}
\end{align}
with the two last terms of~\eqref{eq:app2:h_theta_closed_form1} vanishing as $h_{\theta}(r<R_{\text{i}})=0$ and $r^* \rightarrow R_{\text{i}}$ in the weak penetration asymptotic regime (with the integrand $\partial_{\theta} \dot{h}_r$ being bounded). The integral of~\eqref{eq:app2:faraday} is
\begin{equation}
    e_z(r^*,\theta) = \int_{R_{\text{i}}^-}^{r^*} \mu_0 j_{\text{c}}\,\dot{R}_{\text{i}}\,dr = \mu_0 j_{\text{c}}\,\dot{R}_{\text{i}}\,(r^*-R_{\text{i}}). \label{eq:app2:e_z}
\end{equation}
Inserting~\eqref{eq:app2:j_z} and~\eqref{eq:app2:e_z} into~\eqref{eq:app2:powerLoss1} yields:
\begin{align}
    q_{\text{hys},1} &= - \frac{8}{\pi d_{\text{f}}^2} \mu_0 j_{\text{c}}^2 \int_0^{\pi} \dot{R}_{\text{i}}\, \int_{R_{\text{i}}}^{d_{\text{f}}/2} (r-R_{\text{i}})\,\overbrace{r}^{\approx d_{\text{f}}/2}\,drd\theta \nonumber \\
    &= - \frac{2 \mu_0 j_{\text{c}}^2}{\pi d_{\text{f}}} \int_0^{\pi} \dot{R}_{\text{i}}\,\left( d_{\text{f}}/2 - R_{\text{i}} \right)^2\,d\theta.
\end{align}
In the CSM, a surface current $K$ (A/m) distribution of $|K_z| = 2b_{\text{e}}/\mu_0 |\sin\theta|$ on the boundary of the filament shields from an applied flux density $b_{\text{e}}$~\cite{Carr}. In the weak penetration approximation, the surface current can be approximated by $|K_z| = j_{\text{c}} \cdot (d_{\text{f}}/2 - R_{\text{i}})$, leading to:
\begin{equation}
    d_{\text{f}}/2 - R_{\text{i}} = \frac{2b_{\text{e}}}{j_{\text{c}} \mu_0}|\sin\theta| \quad \Rightarrow \quad \dot{R}_{\text{i}} = -\frac{2}{j_{\text{c}} \mu_0}\dot{b}_{\text{e}}\,|\sin\theta|
\end{equation}
\vspace*{-1em}
\begin{align}
    \Rightarrow q_{\text{hys},1} &= \frac{16}{\pi d_{\text{f}} j_{\text{c}} \mu_0^2} b_{\text{e}}^2~\dot{b}_{\text{e}} \int_0^{\pi} \left(\sin\theta\right)^3\,d\theta \nonumber \\ &= \frac{64}{3\pi d_{\text{f}} j_{\text{c}} \mu_0^2} b_{\text{e}}^2~\dot{b}_{\text{e}}.
\end{align}
}

\section*{Acknowledgment}
\noindent The authors would like to thank W. Kleeven from IBA, M. Wozniak and J. Dular from CERN, S. Schöps from TU Darmstadt and D. Colignon from the University of Liège for fruitful discussions. Computational resources are provided by the Consortium des \'Equipements de Calcul Intensif (C\'ECI), funded by the Fonds de la Recherche Scientifique de Belgique (F.R.S.-FNRS) under Grant No. 2.5020.11.



\end{document}